\documentclass[%
reprint,
superscriptaddress,
 amsmath,amssymb,
 aps,
]{revtex4-2}

\usepackage{graphicx}
\usepackage{dcolumn}
\usepackage{bm}
\usepackage{hyperref}
\usepackage{tikz-cd}
\usepackage[compat=1.1.0]{tikz-feynman}
\usepackage{physics}

\usepackage{cleveref}
\crefname{equation}{Eq.}{Eqs.}

\allowdisplaybreaks
\hypersetup{breaklinks=true}

\begin{document}

\preprint{APS/123-QED}

\title{Effective Field Theories on the Jet Bundle}

\author{Nathaniel Craig}
\affiliation{%
 Department of Physics, University of California, Santa Barbara, California 93106, USA
}%
\affiliation{%
 Kavli Institute for Theoretical Physics, Santa Barbara, California 93106, USA
}
\author{Yu-Tse Lee}%
\affiliation{%
 Department of Physics, University of California, Santa Barbara, California 93106, USA
}%


\begin{abstract}
We develop a generalized field space geometry for higher-derivative scalar field theories, expressing scattering amplitudes in terms of a covariant geometry on the all-order jet bundle. The incorporation of spacetime and field derivative coordinates solves complications due to higher-order derivatives faced by existing approaches to field space geometry. We identify a jet bundle analog to the field space metric that, besides field redefinitions, exhibits invariance under total derivatives. The invariance consequently extends to its amplitude contributions and the canonical covariant geometry.
\end{abstract}

\maketitle


\section{Introduction}

Scattering amplitudes in quantum field theory are invariant under field redefinitions \cite{Chisholm:1961tha, Kamefuchi:1961sb, Arzt:1993gz}, a property aptly framed in terms of coordinate independence in differential geometry. Indeed, it is well-understood that the field space manifold, endowed with a Riemannian metric, identifies amplitudes with covariant tensors like its curvature \cite{Dixon:1989fj, Alonso:2015fsp}. This formalism finds many uses in the study of effective field theories \cite{Alonso:2016oah, Nagai:2019tgi, Helset:2020yio, Cohen:2020xca, Cohen:2021ucp, Alonso:2021rac}, but suffers a significant shortcoming: the Riemannian geometry is unable to naturally accommodate operators involving higher-order derivatives on fields, so that prior manipulation of the Lagrangian is necessary to fit such operators into the Riemannian framework.

To be concrete, consider a theory of scalar fields $\phi^i(x^\mu)$ on the spacetime manifold $\mathcal{T}$ endowed with the Minkowski metric $\eta_{\mu\nu} = \text{diag}(+,-,-,-)$. Just as $x^\mu$ charts out $\mathcal{T}$, $\phi^i$ charts out a manifold $\mathcal{M}$, the field space manifold, with dimension equal to the number of flavors. Field redefinitions $\phi^i(\tilde{\phi}^j)$ without derivatives are then coordinate transformations on $\mathcal{M}$, under which first field derivatives $\partial_\mu \phi^i \equiv \phi^i_\mu$ transform like a tensor on $\mathcal{M}$:
\begin{equation}
    \phi^i_\mu = \frac{\partial \phi^i}{\partial \tilde{\phi}^j} \tilde{\phi}^j_\mu \, .
\end{equation}
Thus, given a Lagrangian $\mathcal{L}$ comprising
\begin{equation}
\label{eq:4derivL}
    V(\phi^a) + \eta^{\mu\nu} g_{ij}(\phi^a) \, \phi^i_\mu \phi^j_\nu + \eta^{\mu\nu} \eta^{\rho\sigma} c_{ijkl}(\phi^a) \, \phi^i_\mu \phi^j_\nu \phi^k_\rho \phi^l_\sigma ,
\end{equation}
the coefficients $V$, $g_{ij}$, and $c_{ijkl}$ transform as tensors. In particular, $g_{ij}$ is positive-definite in a unitary theory and hence a Riemannian metric, giving rise to a Levi-Civita connection $\nabla$ on $\mathcal{M}$ with an associated curvature tensor $R^i_{\;jkl}$. Scattering amplitudes can then be expressed using covariant derivatives of these tensors under $\nabla$, making field redefinition invariance manifest \cite{Cohen:2021ucp}. However, consider a term involving $\partial_\mu \partial_\nu \phi^i \equiv \phi^i_{\mu\nu}$:
\begin{equation}
    \mathcal{L} \supset \eta^{\mu\nu} h_i(\phi^a) \, \phi^i_{\mu\nu} \, .
\end{equation}
Second field derivatives do not transform like tensors:
\begin{equation}
    \phi^i_{\mu_1 \mu_2} = \frac{\partial^2 \phi^i}{\partial \tilde{\phi}^j \partial \tilde{\phi}^k} \tilde{\phi}^j_{\mu_1} \tilde{\phi}^k_{\mu_2} + \frac{\partial \phi^i}{\partial \tilde{\phi}^j} \tilde{\phi}^j_{\mu_1 \mu_2} \, ,
\end{equation}
a complication that makes $g_{ij}$ a tensor no longer:
\begin{equation}
\label{eq:ghtrans}
    \tilde{g}_{kl} = \frac{\partial \phi^i}{\partial \tilde{\phi}^k} \frac{\partial \phi^j}{\partial \tilde{\phi}^l} g_{ij} + \frac{\partial^2 \phi^i}{\partial \tilde{\phi}^k \partial \tilde{\phi}^l} h_i, \quad \tilde{h}_k = \frac{\partial \phi^i}{\partial \tilde{\phi}^k} h_i \, ,
\end{equation}
and derails the path to covariant amplitudes \footnote{Such terms can first be treated using integration by parts or derivative field redefinitions before covariantization, but these manipulations change the apparent field space geometry.}.

To understand why covariance on $\mathcal{M}$ is too limited to treat a general Lagrangian, recall that the tangent space at a point $\phi^i \in \mathcal{M}$ consists of vectors $\varphi^i$ indicating the directions in which one can tangentially pass through $\phi^i$. Precisely speaking, $\varphi^i$ represents an equivalence class of tangent curves $\phi'^i(t)$ with $\phi'^i(0) = \phi^i$, whose first derivatives $d\phi'^i/dt |_{t=0}$ equal $\varphi^i$. Cotangent spaces are the dual spaces of tangent ones, and tensors are built from products of tangent and cotangent spaces at each point. Observe that this construction is based on one-dimensional tangent curves and not $(\dim \mathcal{T})$-dimensional fields. When identifying tensors like $V$ and $c_{ijkl}$, we are abusing the fact that $\phi^i_\mu$ transforms identically to $\varphi^i$. Neither $\phi^i_\mu$ nor $\mathcal{L}$ nor any of its summands is a tensor on $\mathcal{M}$, even if their transformation laws resemble one. Therefore, working on $\mathcal{M}$ only, we are forced to pick out by hand its covariant quantities within $\mathcal{L}$. A more principled approach is to make the whole Lagrangian a covariant object living on some larger manifold, and use that larger covariant geometry to derive scattering amplitudes.

In this Letter, we realize the more principled approach by adding to the field space manifold all degrees of freedom necessary to accommodate spacetime and higher-derivative field coordinates, so that any Lagrangian can be embedded into what are known as {\it jet bundles}. Intuitively, these are bundles over spacetime whose fibers comprise fields and their spacetime derivatives up to a given order \footnote{For a broader overview of the geometry of jet bundles, see, e.g., \cite{saunders1989, olver1986, anderson1992}. For concurrent work applying jet bundles to the geometry of effective field theories, see \cite{Alminawi:2023qtf}.}. Consequently, any scattering amplitude can be written in terms of covariant objects thereon known as distinguished (\textit{d}) tensors. As minimal degrees of freedom are added, the countable dimensionality of this framework offers an advantage over alternative approaches that index fields by position or momentum \cite{Cohen:2022uuw, Cheung:2022vnd}.

After a concise introduction to jet bundles, we construct \textit{d}-tensor derivatives on the first-order jet bundle and use them to obtain covariant expressions for the tree-level scattering amplitudes of the Lagrangian in \cref{eq:4derivL}. We then illustrate the relative advantages of jet bundles at second order, where the resulting geometry automatically encodes invariance under total derivatives, before extending the construction of \textit{d}-tensor derivatives to arbitrary order. Along the way we develop a number of original results, including a geometry covariant under induced fiberwise transformations on an all-order jet bundle with a multidimensional base space, the systematic covariantization of amplitude contributions from any field theory operator, and a geometric interpretation of invariance under total derivatives for the two-derivative part of the theory. Detailed derivations of all-order results, covariant expressions for scattering amplitudes of the most general four-derivative scalar Lagrangian, and the connection with the closely related notion of Lagrange spaces \cite{Craig:2023wni} are discussed in the Supplemental Material \footnote{See Supplemental Material at [url], which includes \cite{Miron:1994nvt,neagu2006,balan2011}, for the construction of a nonlinear connection at arbitrary order, the covariantization of scattering amplitudes for a general four-derivative scalar theory, and the relation between the covariant jet bundle geometry and Lagrange spaces.}. The jet bundle formalism promises to significantly expand the relevance of geometric methods to effective field theories.

\section{Jet Bundles}

We begin with the essential ingredients required to construct covariant amplitudes on the jet bundle.
The first important ingredient is the {\it multi-index}, which suitably collects field derivatives to form derivative coordinates. As partial derivatives commute, field derivatives are uniquely specified by the number of partial derivatives in each spacetime coordinate. To this end, define a multi-index $\Lambda(\tau)$ as a tuple of $(\dim \mathcal{T})$ non-negative integers, where $\tau$ is a spacetime index. The order is given by $|\Lambda| \equiv \sum_\tau \Lambda(\tau)$. Denote
\begin{equation}
    \frac{\partial^{|\Lambda|}}{\partial x^\Lambda} \equiv \prod_\tau \left ( \frac{\partial}{\partial x^\tau} \right )^{\Lambda(\tau)}, \quad \phi^i_\Lambda \equiv \frac{\partial^{|\Lambda|} \phi^i}{\partial x^\Lambda} \, .
\end{equation}
We take $\phi^i_\Lambda$ with $|\Lambda|=0$ to mean $\phi^i$. The subscript $\mu$ in $\phi^i_\mu$ can be interpreted as an order-one multi-index $\Lambda(\tau)$ with one for $\tau = \mu$ and zero otherwise. At higher orders, we may for convenience write $\phi_{\mu_1 \mu_2}$ or $\phi_{\mu_2 \mu_1}$ in place of $\phi_\Lambda$ where $\Lambda(\tau) = 1$ if $\tau = \mu_1 \text{ or } \mu_2$ and 0 otherwise.

A Lagrangian is a function in $\phi^a$ and $\phi^a_\Lambda$. To fit it into an enlargement of $\mathcal{M}$, we need to incorporate derivatives in spacetime of arbitrary order $|\Lambda|$. This can be done through a natural extension of the tangent space construction, which as we have seen in the introduction generates first derivatives in a single time dimension. Let us attach the spacetime manifold to $\mathcal{M}$ to form the trivial fiber bundle $E = \mathcal{T} \times \mathcal{M}$ with an associated projection $\pi: E \rightarrow \mathcal{T}$ \footnote{The trivial bundle suffices since our ultimate interest lies in local rather than global geometry.}. A field $\phi^i(x^\mu)$ is then simply a section of $E$. For $\mathsf{q} \geq 1$, we say that two sections $\phi^i$ and $\phi'^i$ have the same $\mathsf{q}$ jet at a spacetime point $x^\mu \in \mathcal{T}$ if
\begin{equation}
    \left . \frac{\partial^{|\Lambda|} \phi^i}{\partial x^\Lambda} \right |_{x^\mu} = \left . \frac{\partial^{|\Lambda|} \phi'^i}{\partial x^\Lambda} \right |_{x^\mu} \text{ for all } 0 \leq |\Lambda| \leq \mathsf{q} \, ,
\end{equation}
i.e. their derivatives match at $x^\mu$ up to $\mathsf{q}$th order. Having the same $\mathsf{q}$ jet is an equivalence relation between sections, and the set of such equivalence classes over all spacetime points is known as the $\mathsf{q}$th-order jet bundle $J^\mathsf{q}(\pi)$ \cite{ehresmann1953, saunders1989}. This is a manifold charted by $(x^\mu, \phi^i, \phi^i_\Lambda)$ where the derivative coordinates run up to $|\Lambda| \leq \mathsf{q}$, with fiber bundle structures over $\mathcal{T}$, $E$ and $J^\mathsf{r}(\pi)$ for $\mathsf{r} < \mathsf{q}$ given by the projections
\begin{equation}
\begin{gathered}
\begin{cases}
    \pi_\mathsf{q}: J^\mathsf{q}(\pi) \rightarrow \mathcal{T} \\
    (x^\mu, \phi^i, \phi^i_\Lambda) \mapsto (x^\mu)
\end{cases}, \quad
\begin{cases}
    \pi_{\mathsf{q},0}: J^\mathsf{q}(\pi) \rightarrow E \\
    (x^\mu, \phi^i, \phi^i_\Lambda) \mapsto  (x^\mu, \phi^i)
\end{cases}, \\
\begin{cases}
    \pi_{\mathsf{q},\mathsf{r}}: J^\mathsf{q}(\pi) \rightarrow J^\mathsf{r}(\pi) \\
    (x^\mu, \phi^i, \phi^i_\Lambda) \mapsto  (x^\mu, \phi^i, \phi^i_\Theta)
\end{cases} ,
\end{gathered}
\end{equation}
where $|\Theta| \leq \mathsf{r}$. Having enlarged $\mathcal{M}$ to $J^\mathsf{q}(\pi)$, it is now possible to embed any Lagrangian with $\mathsf{q}$th-or-lower field derivatives---it is simply a function $\mathcal{L}: J^\mathsf{q}(\pi) \rightarrow \mathbb{R}$.

The second key ingredient is the notion of a \textit{d} tensor. For our purposes, a \textit{d} tensor is a jet bundle object $T^{\mu i \ldots}_{j \nu \ldots}(x^\alpha, \phi^a, \phi^a_\Lambda)$ that transforms under a field redefinition $\phi^i(\tilde{\phi}^j)$ as if it were a tensor on $\mathcal{M}$ \cite{neagu2005}:
\begin{equation}
    \tilde{T}^{\mu i \ldots}_{j \nu \ldots} = \left ( \frac{\partial \tilde{\phi}^i}{\partial \phi^k} \ldots \right ) T^{\mu k \ldots}_{l \nu \ldots} \left ( \frac{\partial \phi^l}{\partial \tilde{\phi}^j} \ldots \right ) \, .
\end{equation}
Examples of \textit{d} tensors include $\mathcal{L}$ and $\phi^i_\mu$.

A route to obtaining scattering amplitudes from $\mathcal{L}$ is to organize its pieces by derivative count and read off Feynman rules from the coefficients, e.g. $V$, $g_{ij}$ and $c_{ijkl}$ in \cref{eq:4derivL}. This is equivalent to differentiating $\mathcal{L}$ by field derivatives and evaluating on the null section $\{\phi^i_\Lambda = 0\}$ of the jet bundle. However, the process must now be carried out covariantly, which requires \textit{d}-tensor derivatives.

Let us begin with the simplest case $\mathsf{q}=1$, i.e., the first-order jet bundle. The chain rule induces the following transformations of tangent vectors of field and derivative coordinates:
\begin{equation}
    \frac{\partial}{\partial \tilde{\phi}^i} = \frac{\partial \phi^j}{\partial \tilde{\phi}^i} \frac{\partial}{\partial \phi^j} + \frac{\partial \phi^j_\rho}{\partial \tilde{\phi}^i} \frac{\partial}{\partial \phi^j_\rho}, \quad \frac{\partial}{\partial \tilde{\phi}^i_\mu} = \frac{\partial \phi^j}{\partial \tilde{\phi}^i} \frac{\partial}{\partial \phi^j_\mu} \, .
\end{equation}
Evidently, not all are \textit{d} tensors. A nonlinear connection $N$ on $J^1(\pi)$ makes the \textit{d}-tensor combination
\begin{equation}
    \frac{\delta}{\delta \phi^i} \equiv \frac{\partial}{\partial \phi^i} - N^j_{\rho i} \frac{\partial}{\partial \phi^j_\rho} \, .
\end{equation}
The basis $\{\delta/\delta \phi^i, \partial/\partial \phi^i_\mu\}$ spans horizontal (field) and vertical (derivative) distributions in $TJ^1(\pi)$. The coefficient $N^j_{\rho i}$ must transform as
\begin{equation}
\label{eq:N01trans}
    N^k_{\rho l} \frac{\partial \phi^l}{\partial \tilde{\phi}^i} = \tilde{N}^j_{\rho i} \frac{\partial \phi^k}{\partial \tilde{\phi}^j} - \frac{\partial \phi^k_\rho}{\partial \tilde{\phi}^i} = \tilde{N}^j_{\rho i} \frac{\partial \phi^k}{\partial \tilde{\phi}^j} - \frac{\partial^2 \phi^k}{\partial \tilde{\phi}^i \partial \tilde{\phi}^j} \tilde{\phi}^j_\rho \, .
\end{equation}

In the Riemannian formulation of amplitudes on $\mathcal{M}$, covariant derivatives are supplied by the affine connection $\nabla$, whose Christoffel symbols $\gamma^i_{\;jk}(\phi^a)$ transform appropriately to cancel the nontensorial parts of composed partial derivatives. In extending this covariant framework to jet bundles, we are thus motivated to construct $N$ from an affine connection on $\mathcal{M}$. It is not essential that $\nabla$ be Levi-Civita, and we make no assumption other than it be symmetric. Then the following choice suffices \cite{neagu2003}:
\begin{equation}
    N^j_{\rho i} = \gamma^j_{\;ik} \phi^k_\rho \, .
\end{equation}

We have now constructed single \textit{d}-tensor derivatives. But we will need an additional affine connection $\grad$ on $J^1(\pi)$, which is $N$ linear in the sense that its parallel transport respects the horizontal-vertical decomposition by $N$ \cite{neagu2001}:
\begin{equation}
\label{eq:Nlinear}
    \grad_{\frac{\delta}{\delta \phi^k}} \frac{\delta}{\delta \phi^j_\Lambda} = F^i_{\;jk} \frac{\delta}{\delta \phi^i_\Lambda}, \quad \grad_{\frac{\partial}{\partial \phi^k_\rho}} \frac{\partial}{\partial \phi^j_\Lambda} = C^{i \rho}_{\;jk} \frac{\partial}{\partial \phi^i_\Lambda} \, ,
\end{equation}
where $|\Lambda| \geq 0$. The coefficients $F^i_{\;jk}$ and $C^{i\rho}_{\;jk}$ are $h$- and $v$-Christoffel symbols, transforming like Christoffel symbols and tensors on $\mathcal{M}$ respectively. We can choose
\begin{equation}
    F^i_{\;jk} = \gamma^i_{\;jk} , \quad C^{i\rho}_{\;jk} = 0 .
\end{equation}
This yields $h$- and $v$-covariant derivatives on any \textit{d} tensor $T$:
\begin{subequations}
\begin{align}
    &(\grad T)^{i \ldots}_{\;j \ldots} = T^{i \ldots}_{\;j \ldots / k} \, d\phi^k + T^{i \ldots}_{\;j \ldots} |_k^\rho \, \delta \phi^k_\rho , \\
    &\grad_k T^{i \ldots}_{\;j \ldots} \equiv T^{i \ldots}_{\;j \ldots / k} = \frac{\delta T^{i \ldots}_{\;j \ldots}}{\delta \phi^k} + F^i_{\;mk} T^{m \ldots}_{\;j \ldots} - F^m_{\;jk} T^{i \ldots}_{\;m \ldots} + \ldots , \\
    &\grad_k^\rho T^{i \ldots}_{\;j \ldots} \equiv T^{i \ldots}_{\;j \ldots} |_k^\rho = \frac{\partial T^{i \ldots}_{\;j \ldots}}{\partial \phi^k_\rho},
\end{align}
\end{subequations}
allowing us to compose \textit{d}-tensor derivatives.

Equipped with the \textit{d} tensor $\mathcal{L}$ and a covariant way to differentiate it on $J^1(\pi)$, we can proceed to assemble covariant amplitudes.

\section{Covariant Amplitudes}

Consider the Lagrangian in \cref{eq:4derivL}. We call $\bar{\phi}^i = \arg \min V(\phi^a)$ and $\{\phi^i = \bar{\phi}^i, \phi^i_\mu = 0 \}$ the vacuum on $\mathcal{M}$ and $J^1(\pi)$, respectively; the latter lies on the null section of $J^1(\pi)$. Evaluation of geometric quantities in either space at the corresponding vacuum will be denoted with an overline. To compute tree-level amplitudes, we expand $V$, $g_{ij}$, and $c_{ijkl}$ about $\bar{\phi}^i$, indicating partial derivatives with commas. In coordinates that diagonalize $\bar{V}_{,ij}$ and $\bar{g}_{ij}$, the scalar field masses can be read off from their ratios $\bar{V}_{,ij} = -2 \bar{g}_{ij} m_i^2$. The Feynman rules in momentum space are then
\begin{subequations}
\begin{alignat}{2}
    \vcenter{\hbox{    \begin{tikzpicture} \begin{feynman}
        \vertex (a) at (-0.6,0) {$i$};
        \vertex (b) at (0.6,0) {$j$};
        \diagram* {(a) -- (b)};
    \end{feynman} \end{tikzpicture}}}
    \> &= \>
    &&\frac{i\bar{g}^{ij}}{2(p^2 - m_i^2)} \, , \\
    \vcenter{\hbox{ \begin{tikzpicture} \begin{feynman}
            \vertex (m) at (0,0);
            \vertex (a1) at (0.9,0) {$i_1$};
            \vertex (a2) at (0.8,0.5) {$i_2$};
            \vertex (ar) at (0.8,-0.5) {$i_\mathtt{n}$};
            \diagram* {
            (a1) -- (m) -- (a2),
            (ar) -- (m)};
            \draw [dotted,thick,domain=45:315,scale=0.5] plot ({cos(\x)}, {sin(\x)});
    \end{feynman} \end{tikzpicture}}}
    &= \> &&i \left [ \bar{V}_{, \,\ldots} - 2 \sum_{1 \leq \mathtt{a} < \mathtt{b} \leq \mathtt{n}} (p_\mathtt{a} \cdot p_\mathtt{b}) \bar{g}_{i_\mathtt{a} i_\mathtt{b}, \ldots} \right. \\
    & &&\quad \left. + \, 8 \sum_{\mathtt{a} < \mathtt{b}, \mathtt{a} < \mathtt{c} < \mathtt{d}} (p_\mathtt{a} \cdot p_\mathtt{b}) (p_\mathtt{c} \cdot p_\mathtt{d}) \bar{c}_{i_\mathtt{a} i_\mathtt{b} i_\mathtt{c} i_\mathtt{d}, \ldots} \right ] \, , \nonumber
\end{alignat}
\end{subequations}
where the ellipses represent indices from $i_1$ to $i_\mathtt{n}$ that are not explicitly written.

At tree level, any momentum $p$ appearing in a Feynman diagram can be traded for Mandelstam variables, which are external kinematic data that can be set aside. Rewriting any $m_i^2$ as $- \bar{g}^{ii} \bar{V}_{,ii} / 2$, all that remain in the amplitude are partial derivatives of $V$, $g$, or $c$ evaluated at the vacuum, which we desire to convert to covariant expressions.

To do so, we deploy the following trick. In the normal coordinates of a symmetric $\nabla$ at the vacuum on $\mathcal{M}$, one can replace partial derivatives on any tensor $t^{i \ldots}_{\;j \ldots}(\phi^a)$ with covariant ones \cite{Cohen:2021ucp, Craig:2023wni}:
\begin{equation}
\label{eq:normalreplace}
    \overline{t}^{i \ldots}_{\;j \ldots \, , \, k_1 \ldots k_\mathtt{n}} \rightarrow \nabla_{(k_1} \ldots \nabla_{k_\mathtt{n})} \overline{t}^{i \ldots}_{\;j \ldots} + \mathcal{O}(tR) \, ,
\end{equation}
incurring additional covariant terms that involve the curvature tensor $R^i_{\;jkl}$ of $\nabla$, suppressed above. Now returning to $J^1(\pi)$, it is also easy to verify that
\begin{equation}
\label{eq:ttoT}
    \left ( \grad_{k_1} \ldots \grad_{k_\mathtt{n}} T^{i \ldots}_{j \ldots} \right ) \Bigr |_{\phi^a_\alpha = 0} = \nabla_{k_1} \ldots \nabla_{k_\mathtt{n}} \left ( T^{i \ldots}_{j \ldots} \Bigr |_{\phi^a_\alpha = 0} \right ) \, ,
\end{equation}
for any \textit{d} tensor $T(\phi^a, \phi^a_\alpha)$. We can hence replace $t$ in \cref{eq:normalreplace} with any such $T$ that agrees with $t$ on the null section, $\nabla$ with $\grad$, and $R$ with the $hh$-curvature \textit{d} tensor $\mathcal{R}$ of $\grad$ \footnote{All other curvature components of $\grad$ vanish identically.}:
\begin{align}
    \mathcal{R}^i_{\;jkl} &\equiv d\phi^i \left ( \left ( \left [\grad_{\delta / \delta \phi^k} , \grad_{\delta / \delta \phi^l} \right ] - \grad_{[\delta / \delta \phi^k, \delta / \delta \phi^l]} \right ) \frac{\delta}{\delta \phi^j} \right ) \nonumber \\
    &= \frac{\delta \gamma^i_{\;jl}}{\delta \phi^k} + \gamma^i_{\;mk} \gamma^m_{\;jl} - (k \leftrightarrow l) = R^i_{\;jkl} \, .
\end{align}
The result is a manifestly covariant expression at the vacuum of $J^1(\pi)$ that, in normal coordinates, equals the desired partial derivative of $t(\phi^a)$ at the vacuum of $\mathcal{M}$.

For a $\mathsf{q}=1$ Lagrangian, the coefficients on $\mathcal{M}$ that appear in the Feynman rules are always tensors, and their \textit{d}-tensor counterparts are simply partial derivatives of $\mathcal{L}$ by $\phi^i_\mu$. For example,
\begin{gather}
\label{eq:Vgcreplace}
    V = \mathcal{L} \, \Bigr |_{\phi^a_\alpha = 0}, \quad g_{ij} = \frac{\eta_{\mu\nu}}{8} \frac{\partial^2 \mathcal{L}}{\partial \phi^i_\mu \partial \phi^j_\nu} \Bigr |_{\phi^a_\alpha = 0}, \\
    c_{ijkl} = \frac{5 \eta_{\mu\nu} \eta_{\rho\sigma} - \eta_{\mu\rho} \eta_{\nu\sigma} - \eta_{\mu\sigma} \eta_{\nu\rho}}{576} \frac{\partial^4 \mathcal{L}}{\partial \phi^i_\mu \partial \phi^j_\nu \partial \phi^k_\rho \partial \phi^l_\sigma} \Bigr |_{\phi^a_\alpha = 0} \, . \nonumber
\end{gather}
We can thus use the trick to write all nonkinematic terms in the scattering amplitude as \textit{d} tensors in normal coordinates. But since the total amplitude is covariant, the ensuing covariant expression must actually hold in any coordinates, and we are done.

This procedure is largely the same as in \cite{Craig:2023wni}. Denoting the $n$-point amplitude as $\mathcal{A}_{1 \ldots \mathtt{n}}$ where $1, \ldots, \mathtt{n}$ label the external legs, we find
\begin{widetext}
\begin{subequations}
\label{eq:amplitudes}
\begin{align}
    &\mathcal{A}_{123} \>\, = \prod_{\mathtt{e}=1}^3 \sqrt{\frac{\bar{g}^{\mathtt{e}\mathtt{e}}}{2}} \left \{ \bar{V}_{/(123)} - \frac{1}{2} \sum_{\mathtt{a}=1}^3 \bar{V}_{/\mathtt{a}j} \, \bar{g}^{jk} \left( 2 \bar{g}_{k(\mathtt{b}/\mathtt{c})} - \bar{g}_{\mathtt{b}\mathtt{c}/k} \right) \right \} \, , \\
    &\mathcal{A}_{1234} = \prod_{\mathtt{e}=1}^4 \sqrt{\frac{\bar{g}^{\mathtt{e}\mathtt{e}}}{2}} \left \{ \left [ - \frac{1}{2} A_{12j} \frac{\bar{g}^{jk}}{s_{12} - m_j^2} A_{34k} + \text{($s_{13}$, $s_{14}$ channels)} \right ] + \bar{V}_{/(1234)} + \sum_{\mathtt{a}=1}^4 m_\mathtt{a}^2 \left [3\, \bar{g}_{\mathtt{a}(\mathtt{b}/\mathtt{c}\mathtt{d})} + \bar{R}_{(\mathtt{b}\mathtt{c}\mathtt{d})\mathtt{a}} \right ] \right. \\
    &\hspace{3.3em} \left. + \, 4 \sum_{\mathtt{a}<\mathtt{b}} m_\mathtt{a}^2 m_\mathtt{b}^2 \, \bar{c}_{1234} - \sum_{\mathtt{a}<\mathtt{b}} s_{\mathtt{a}\mathtt{b}} \left[ \bar{g}_{\mathtt{a}\mathtt{b}/(\mathtt{c}\mathtt{d})} + \frac{1}{3} \bar{R}_{\mathtt{a}(\mathtt{c}\mathtt{d})\mathtt{b}} + \frac{1}{3} \bar{R}_{\mathtt{b}(\mathtt{c}\mathtt{d})\mathtt{a}} + 2 (m_\mathtt{c}^2 + m_\mathtt{d}^2) \, \bar{c}_{1234} \right] + 2 (s_{12}^2 + s_{13}^2 + s_{14}^2) \, \bar{c}_{1234} \right \} \nonumber \\
    &\text{where } A_{\mathtt{a} \mathtt{b} i} = \bar{V}_{/(\mathtt{a} \mathtt{b} i)} - \frac{1}{2} \bar{V}_{/\mathtt{a}j} \, \bar{g}^{jk} \left( 2 \bar{g}_{k(\mathtt{b}/i)} - \bar{g}_{\mathtt{b}i/k} \right) - \frac{1}{2} \bar{V}_{/\mathtt{b}j} \, \bar{g}^{jk} \left( 2 \bar{g}_{k(\mathtt{a}/i)} - \bar{g}_{\mathtt{a}i/k} \right) + s_{\mathtt{a} \mathtt{b}} \left( 2 \bar{g}_{i(\mathtt{a}/\mathtt{b})} - \bar{g}_{\mathtt{a}\mathtt{b}/i} \right) \, , \nonumber
\end{align}
\end{subequations}
\end{widetext}
and so on. For brevity, apart from $\mathcal{A}$, $A$, and Mandelstam variables $s_{\mathtt{a}\mathtt{b}} = (p_\mathtt{a} + p_\mathtt{b})^2$, any $\mathtt{a}$ or explicit number appearing in a subscript should be interpreted as flavor index $i_\mathtt{a}$, and any tensor on $\mathcal{M}$ as its \textit{d}-tensor replacement in \cref{eq:Vgcreplace}. The indices $\mathtt{a}$, $\mathtt{b}$, $\mathtt{c}$, and $\mathtt{d}$ are all distinct, and the $\sqrt{\bar{g}^{\mathtt{e}\mathtt{e}}/2}$ are wave function normalization factors.

Of course, instead of the covariant geometry of $\grad$ on the null section of $J^1(\pi)$, that of $\nabla$ on $\mathcal{M}$ would have sufficed to covariantize scattering amplitudes if we terminate the replacement procedure at \cref{eq:normalreplace}. However, since $\mathcal{L}$ does not fit on $\mathcal{M}$, we then have to endow $\mathcal{M}$ with tensors $V$, $g$, and $c$ manually extracted from $\mathcal{L}$. By enlarging the manifold, we have eliminated the need for additional structures besides $\mathcal{L}$ and $\grad$ in expressing amplitudes covariantly, even if intermediate steps might involve other temporary objects. Moreover, the covariant geometry of the order-one jet bundle paves the way to higher field derivatives, which field space alone cannot handle systematically. This brings us to the next order---$\mathsf{q} = 2$.

\section{Invariance under Total Derivatives}

The construction of \textit{d}-tensor derivatives on the second-order jet bundle proceeds in close analogy with the first-order case. The tangent vectors of $J^2(\pi)$ transform as \cite{oana2018}\footnote{There are two summation conventions in effect here. There are $\binom{\dim \mathcal{T} + \mathsf{q} - 1}{\mathsf{q}}$ multi-indices of order $\mathsf{q}$ but $(\dim \mathcal{T})^\mathsf{q}$ subscripts. If we repeat a subscript styled with the same Greek letter distinctly numbered, e.g. $\mu_1 \mu_2$, we mean a sum over distinct multi-indices, not independent summations over each index $\mu_1$ and $\mu_2$.}
\begin{equation}
\begin{gathered}
    \frac{\partial}{\partial \tilde{\phi}^i} = \frac{\partial \phi^j}{\partial \tilde{\phi}^i} \frac{\partial}{\partial \phi^j} + \frac{\partial \phi^j_\rho}{\partial \tilde{\phi}^i} \frac{\partial}{\partial \phi^j_\rho} + \frac{\partial \phi^j_{\rho_1 \rho_2}}{\partial \tilde{\phi}^i} \frac{\partial}{\partial \phi^j_{\rho_1 \rho_2}} \, , \\
    \frac{\partial}{\partial \tilde{\phi}^i_\mu} = \frac{\partial \phi^j}{\partial \tilde{\phi}^i} \frac{\partial}{\partial \phi^j_\mu} + \frac{\partial \phi^j_{\rho_1 \rho_2}}{\partial \tilde{\phi}^i_\mu} \frac{\partial}{\partial \phi^j_{\rho_1 \rho_2}} , \frac{\partial}{\partial \tilde{\phi}^i_{\mu_1 \mu_2}} = \frac{\partial \phi^j}{\partial \tilde{\phi}^i} \frac{\partial}{\partial \phi^j_{\mu_1 \mu_2}} \, .
\end{gathered}
\end{equation}
We require a suitable nonlinear connection $N$ that enables us to combine them into \textit{d} tensors
\begin{subequations}
\begin{align}
    \frac{\delta}{\delta \phi^i} &= \frac{\partial}{\partial \phi^i} - N^j_{\rho i} \frac{\partial}{\partial \phi^j_\rho} - N^j_{\rho_1 \rho_2 i} \frac{\partial}{\partial \phi^j_{\rho_1 \rho_2}} \, , \\
    \frac{\delta}{\delta \phi^i_\mu} &= \frac{\partial}{\partial \phi^i_\mu} - N^{j \mu}_{\rho_1 \rho_2 i} \frac{\partial}{\partial \phi^j_{\rho_1 \rho_2}} \, .
\end{align}
\end{subequations}
The coefficients are all denoted by $N$, but can be distinguished by index structure. The answer, as will be apparent later, is to set
\begin{equation}
\begin{gathered}
    N^j_{\rho i} = \gamma^j_{\;ik} \phi^k_\rho, \quad N^{j \mu}_{\rho_1 \rho_2 i} = 2 N^j_{(\rho_1|i} \delta^\mu_{|\rho_2)} = 2 \gamma^j_{\;ik} \phi^k_{(\rho_1} \delta^\mu_{\rho_2)} \, , \\
    N^j_{\rho_1 \rho_2 i} = \left ( \gamma^j_{\;ik,l} - \gamma^j_{\;mk} \gamma^m_{\;il} \right ) \phi^k_{(\rho_1} \phi^l_{\rho_2)} + \gamma^j_{\;ik} \phi^k_{\rho_1 \rho_2} \, .
\end{gathered}
\end{equation}
An $N$-linear connection then follows from a simple modification of \cref{eq:Nlinear}, with $F^i_{\;jk} = \gamma^i_{\;jk}$ and $C^{i\Theta}_{\;jk} = 0$ for $|\Theta| \geq 1$.

A new feature arises at second order---besides field redefinitions, the Lagrangian can be modified by a total derivative without changing the resulting amplitudes. To illustrate this, it suffices to examine a minimal example, the most general two-derivative Lagrangian
\begin{equation}
    \mathcal{L} = V(\phi^a) + \eta^{\mu\nu} g_{ij}(\phi^a) \, \phi^i_\mu \phi^j_\nu + \eta^{\mu\nu} h_i(\phi^a) \, \phi^i_{\mu\nu} \, ,
\end{equation}
in which integration by parts shuffles $g_{ij}$ and $h_i$. The two pieces really represent a single kinetic term
\begin{equation}
    G_{ij} = g_{ij} - h_{(i,j)} \, ,
\end{equation}
which should be positive-definite. We can follow the same procedure as before to covariantize amplitudes, once we figure out the tensors $t(\phi^a)$ in the Feynman rules to be replaced with \textit{d} tensors $T(\phi^a, \phi^a_\Lambda)$ obtained from $\mathcal{L}$, invoking \cref{eq:ttoT} with $\alpha \rightarrow \Lambda$. Picking out $V$ and $h_i$ from $\mathcal{L}$ is simple because they are indeed tensors,
\begin{equation}
    \mathcal{L} \Bigr |_{\phi^a_\Lambda = 0} = V, \quad \frac{1}{4} \eta_{\mu\nu} \frac{\delta \mathcal{L}}{\delta \phi^i_{\mu\nu}} \Bigr |_{\phi^a_\Lambda = 0} = h_i \, ,
\end{equation}
but since $g_{ij}$ is no longer a tensor, the naive guess for $g_{ij}$ yields an extraneous term,
\begin{equation}
    \frac{1}{8} \eta_{\mu\nu} \frac{\delta^2 \mathcal{L}}{\delta \phi^{(i}_\mu \delta \phi^{j)}_\nu} \Bigr |_{\phi^a_\Lambda = 0} = g_{ij} - \gamma^k_{\;ij} h_k \, .
\end{equation}
The nontensorial term in the transformation law for $\gamma^k_{\;ij}$ precisely cancels the nontensorial term for $g_{ij}$ in \cref{eq:ghtrans}. Eliminating the extraneous term, we should take the difference with the covariant derivative of $h_i$, but that simply returns the \textit{d} tensor for $G_{ij}$:
\begin{equation}
\label{eq:Gdtensor}
    \frac{1}{4} \eta_{\mu\nu} \left ( \frac{1}{2} \frac{\delta^2 \mathcal{L}}{\delta \phi^{(i}_\mu \delta \phi^{j)}_\nu} - \grad_{(i} \frac{\delta \mathcal{L}}{\delta \phi^{j)}_{\mu\nu}} \right ) \Bigg |_{\phi^a_\Lambda = 0} = G_{ij} \, .
\end{equation}
Evidently, the jet bundle geometry automatically identifies a single kinetic term even if we start with two pieces. The amplitudes thence derived from \cref{eq:Gdtensor} exhibit explicit invariance under both field redefinitions and total derivatives.

Thus far, we have remained agnostic on the choice of field space connection that generates the jet bundle connections. But now, there is a natural candidate: the Levi-Civita connection of the kinetic term \cref{eq:Gdtensor}. This choice is canonical as it follows solely from the Lagrangian, independent of the arbitrary $\gamma$ used to derive it. It is also meaningful due to its significance in the Lagrangian, so that any covariant physics arising from the kinetic term is encapsulated by the $hh$ curvature \footnote{Any covariant derivative of the kinetic term in covariantized amplitudes like \cref{eq:amplitudes} will vanish.}. Such a choice is the covariant generalization of that in \cite{Finn:2020nvn} and agrees with the established Riemannian framework when the Lagrangian is actually first-order. Despite now tying the jet bundle connection to a Lagrangian, the resultant covariant geometry will nevertheless remain invariant under total derivatives, as it should be for a framework that extracts local physics.

\section{To All Derivative Orders}

The last obstacle is extending the covariant geometry of the jet bundle to arbitrary order. This can be done using a generalization of prolongation on higher-order tangent bundles \cite{miron1993, miron1996}, with complications due to multi-indices in $\dim \mathcal{T} > 1$. Here, we state the results, leaving a proof to the Supplemental Material \cite{Note3}.

Let $\mathsf{q} \geq \mathsf{r} \geq 0$ and abbreviate the order-$\mathsf{q}$ subscript $\mu_1 \ldots \mu_\mathsf{q}$ as $\mu_\mathsf{Q}$ with an uppercase $\mathsf{Q}$. By $\mu_{\mathsf{Q}-1}$, we mean $\mu_1 \ldots \mu_{\mathsf{q}-1}$. Also write $\delta^{\rho_\mathsf{Q}}_{\mu_\mathsf{Q}}$ for the Kronecker delta that indicates multi-index equality $\rho_\mathsf{Q} = \mu_\mathsf{Q}$. Define
\begin{equation}
    \Gamma_{\mu_0} = \phi^i_{\mu_0} \frac{\partial}{\partial \phi^i} + \phi^i_{\mu_0 \mu_1} \frac{\partial}{\partial \phi^i_{\mu_1}} + \phi^i_{\mu_0 \mu_1 \mu_2} \frac{\partial}{\partial \phi^i_{\mu_1 \mu_2}} + \ldots \, .
\end{equation}
We can assemble \textit{d}-tensor derivatives
\begin{equation}
    \frac{\delta}{\delta \phi^i_{\mu_\mathsf{R}}} = \frac{\partial}{\partial \phi^i_{\mu_\mathsf{R}}} - N^{j \mu_\mathsf{R}}_{\rho_{\mathsf{R}+1} i} \frac{\partial}{\partial \phi^j_{\rho_{\mathsf{R}+1}}} - N^{j \mu_\mathsf{R}}_{\rho_{\mathsf{R}+2} i} \frac{\partial}{\partial \phi^j_{\rho_{\mathsf{R}+2}}} - \ldots \, ,
\end{equation}
by recursively setting the coefficients in the dual basis of \textit{d}-tensor 1-forms as
\begin{equation}
\begin{gathered}
    M^i_{\mu j} = \gamma^i_{\;jk} \phi^k_\mu, \quad M^{i \rho_{\mathsf{Q}-1}}_{\mu_\mathsf{Q} j} = \mathsf{q} M^i_{(\mu_\mathsf{q}|j} \delta^{\rho_{\mathsf{Q}-1}}_{|\mu_{\mathsf{Q} - 1})} \, , \\
    M^{i \rho_\mathsf{R}}_{\mu_\mathsf{Q} j} = \frac{\mathsf{q}}{\mathsf{q}-\mathsf{r}} \left [ \Gamma_{(\mu_\mathsf{q}} M^{i \rho_\mathsf{R}}_{\mu_{\mathsf{Q}-1}) j} + M^i_{(\mu_\mathsf{q}| m} M^{m \rho_\mathsf{R}}_{|\mu_{\mathsf{Q}-1}) j} \right ] \, ,
\end{gathered}
\end{equation}
fixing $N$ by duality
\begin{equation}
    N^{i \rho_\mathsf{R}}_{\mu_\mathsf{Q} j} = M^{i \rho_\mathsf{R}}_{\mu_\mathsf{Q} j} - M^{i \sigma_{\mathsf{Q}-1}}_{\mu_\mathsf{Q} k} N^{k \rho_\mathsf{R}}_{\sigma_{\mathsf{Q}-1} j} - \ldots - M^{i \sigma_{\mathsf{R}+1}}_{\mu_\mathsf{Q} k} N^{k \rho_\mathsf{R}}_{\sigma_{\mathsf{R}+1} j} \, .
\end{equation}
An $N$-linear connection is then given by
\begin{equation}
\label{eq:allorderNlinear}
    \grad_{\frac{\delta}{\delta \phi^k}} \frac{\delta}{\delta \phi^j_\Lambda} = \gamma^i_{\;jk} \frac{\delta}{\delta \phi^i_\Lambda}, \quad \grad_{\frac{\delta}{\delta \phi^k_\Theta}} \frac{\delta}{\delta \phi^j_\Lambda} = 0 \, .
\end{equation}
where $|\Lambda| \geq 0$ and $|\Theta| \geq 1$. With such technology, we can now extract information at any derivative order from any Lagrangian covariantly, and from there derive scattering amplitudes as before. As an example, the fully general four-derivative Lagrangian is treated in the Supplemental Material \cite{Note3}.

\section{Outlook}

Geometry is a potentially powerful organizing principle for diverse effective field theories, provided it can incorporate the full scope of interactions. The jet bundle generalizes the field space manifold by incorporating spacetime and field derivative coordinates of all orders. It therefore contains all degrees of freedom necessary to accommodate any scalar field theory. In this Letter, we constructed a geometry on the jet bundle that transforms covariantly, and demonstrated how the invariance of scattering amplitudes under nonderivative field redefinitions can be made manifest in terms of \textit{d} tensors. In the process, we learned that an important covariant object in the Lagrangian, the kinetic term, also exhibits manifest invariance under total derivatives, yielding a new geometric perspective on amplitude physics and a canonical choice for the covariant geometry.

The establishment of a geometric framework to all derivative orders opens up the possibility of redefinitions mixing field and derivative coordinates, which nevertheless leave amplitudes invariant. Such redefinitions are relevant in practice by virtue of equation-of-motion reduction, often used to eliminate derivative operators. While the current approach produces covariant amplitudes by considering each operator in turn, derivative field redefinitions shuffle contributions between operators so that the sum remains unchanged only when momentum conservation and on-shell conditions are imposed. It remains to be seen whether the geometry of the infinite jet bundle is capable of capturing the more general invariance.

The results in this Letter may be plausibly extended to loop-level amplitudes and higher-spin theories. The employment of normal coordinates on the field space manifold to covariantize the renormalization procedure, such as in \cite{Alonso:2022ffe, Helset:2022pde}, can be translated to jet bundles to explicitly render higher-derivative operators. Beyond scalar fields, gauge bosons can be incorporated by fixing a gauge like in \cite{Helset:2022tlf, Finn:2019aip}, or generalizing the notion of jets to principal and associated bundles; fermions may be incorporated via an extension to supermanifolds \cite{Finn:2020nvn,Assi:2023zid,Gattus:2023gep}.

\section*{Acknowledgements}

The authors would like to thank John Celestial and Timothy Cohen for valuable discussions, Xiaochuan Lu and Dave Sutherland for valuable discussions and comments on the manuscript; Jacques Distler and Seth Koren for previous invocations of jet bundles for EFTs, and Mohammed Alminawi, Ilaria Brivio, and Joe Davighi for correspondence about their related work \cite{Alminawi:2023qtf}. This work was supported in part by the U.S. Department of Energy under the grant DE-SC0011702 and performed in part at the Kavli Institute for Theoretical Physics, supported by the National Science Foundation under grant no.~NSF PHY-1748958.


\bibliography{arxiv_v2}

\clearpage
\widetext
\setcounter{figure}{0}
\renewcommand\thefigure{S\arabic{figure}} 

\setcounter{equation}{0}
\renewcommand\theequation{S\arabic{equation}} 
 
\section*{Supplemental material}

\subsection*{Constructing a Non-linear Connection at Arbitrary Order}

Here we construct a non-linear connection on a jet bundle of arbitrary order, generalizing the derivation for first- and second-order jet bundles presented in the main text. The general case benefits from adopting streamlined notation and conventions for multi-indices. In the following, we reserve $\mathsf{q} \geq \mathsf{r} \geq \mathsf{s} \geq 0$ for derivative orders, and write $\mu_\mathsf{R}$ with an uppercase $\mathsf{R}$ as a shorthand for the multi-index represented by the subscript $\mu_\mathsf{1} \ldots \mu_\mathsf{r}$ if $\mathsf{r} \geq 1$, or the zero multi-index if $\mathsf{r} = 0$. We also reserve the letters $\mathsf{I}$ and $\mathsf{J}$ for generally non-consecutive subsets of $\{1, \ldots, \mathsf{r}\}$, and write $\mu_\mathsf{I}$ for the subscript $\mu_{\mathsf{I}(1)} \mu_{\mathsf{I}(2)} \ldots$. In this sense, $\mathsf{R}$ really stands for the full set $\{1, \ldots, \mathsf{r}\}$. By $\mathsf{R}-1$, we will mean $\{1, \ldots, \mathsf{r}-1\}$. This notation is useful in describing the transformation law for derivative coordinates under a field redefinition $\phi^p = \phi^p(\tilde{\phi}^l)$:
\begin{equation}
    \phi^p_{\mu_\mathsf{R}} = \sum_{\substack{\cup \mathsf{I}_i = \mathsf{R} \\ \mathsf{I}_i \cap \mathsf{I}_j = \varnothing}} \frac{\partial^n \phi^p}{\partial \tilde{\phi}^{l_1} \ldots \partial \tilde{\phi}^{l_n}} \tilde{\phi}^{l_1}_{\mu_{\mathsf{I}_1}} \ldots \tilde{\phi}^{l_n}_{\mu_{\mathsf{I}_n}} \, .
\end{equation}
Each spacetime derivative $\partial_\mu$ acts on $\phi$ through $\tilde{\phi}$ via the chain rule, and the sum runs over all possible ways to do so --- each summand corresponds to an unordered partition of the $\mathsf{r}$ indices $\mu_1, \ldots, \mu_\mathsf{r}$ to assign to the $n \leq \mathsf{r}$ field coordinates $\tilde{\phi}$. For instance, if $\mathsf{r} = 5$, the partition $\tilde{\phi}^{l_1}_{\mu_1 \mu_2} \tilde{\phi}^{l_2}_{\mu_3 \mu_4} \tilde{\phi}^{l_3}_{\mu_5}$ is the same as $\tilde{\phi}^{l_1}_{\mu_2 \mu_1} \tilde{\phi}^{l_2}_{\mu_3 \mu_4} \tilde{\phi}^{l_3}_{\mu_5}$ and $\tilde{\phi}^{l_1}_{\mu_3 \mu_4} \tilde{\phi}^{l_2}_{\mu_1 \mu_2} \tilde{\phi}^{l_3}_{\mu_5}$, but distinct from $\tilde{\phi}^{l_1}_{\mu_3 \mu_2} \tilde{\phi}^{l_2}_{\mu_1 \mu_4} \tilde{\phi}^{l_3}_{\mu_5}$ even if $\mu_1$ and $\mu_3$ happen to take on the same value. The recursive structure of partitions leads to the identity
\begin{equation}
\begin{aligned}
    \frac{\partial \phi^p_{\mu_\mathsf{R}}}{\partial \tilde{\phi}^l_{\nu_\mathsf{S}}} &= \sum_{\substack{\cup \mathsf{I}_i = \mathsf{R} \\ \mathsf{I}_i \cap \mathsf{I}_j = \varnothing}} \sum_{1 \leq k \leq n} \frac{\partial^n \phi^p}{\partial \tilde{\phi}^{l_1} \ldots \partial \tilde{\phi}^l \ldots \partial \tilde{\phi}^{l_n}} \tilde{\phi}^{l_1}_{\mu_{\mathsf{I}_1}} \ldots \delta^{\nu_\mathsf{S}}_{\mu_{\mathsf{I}_k}} \ldots \tilde{\phi}^{l_n}_{\mu_{\mathsf{I}_n}} \\
    &= \sum_{\mathsf{I} \subseteq \mathsf{R}} \delta^{\nu_\mathsf{S}}_{\mu_\mathsf{I}} \frac{\partial}{\partial \tilde{\phi}^l} \sum_{\substack{\cup \mathsf{J}_i = \mathsf{R} \setminus \mathsf{I} \\ \mathsf{J}_i \cap \mathsf{J}_j = \varnothing}} \frac{\partial^m \phi^p}{\partial \tilde{\phi}^{l_1} \ldots \partial \tilde{\phi}^{l_m}} \tilde{\phi}^{l_1}_{\mu_{\mathsf{J}_1}} \ldots \tilde{\phi}^{l_m}_{\mu_{\mathsf{J}_m}} = \sum_{\mathsf{I} \subseteq \mathsf{R}} \delta^{\nu_\mathsf{S}}_{\mu_\mathsf{I}} \frac{\partial \phi^p_{\mu_\mathsf{R} - \mu_\mathsf{I}}}{\partial \tilde{\phi}^l} \, .
\end{aligned}
\end{equation}
We extended the Kronecker delta to be an indicator function for the equality of multi-indices. Multi-index subtraction is to be performed element-wise, valid only when the result is a tuple of non-negative integers. Taking $\dim \mathcal{T} = 4$ for instance, $(0, 1, 3, 1) - (0, 0, 2, 1) = (0, 1, 1, 0)$ while $(0, 1, 3, 1) - (0, 0, 1, 2)$ is not defined. Multi-index addition is defined analogously.

The summation convention for multi-indices deserves some clarification. There are $\binom{\dim \mathcal{T} + \mathsf{q} - 1}{\mathsf{q}}$ multi-indices of order $\mathsf{q}$ but $(\dim \mathcal{T})^\mathsf{q}$ subscripts; if a subscript like $\mu_1 \mu_2$ is repeated, it will mean an implicit summation over distinct multi-indices, not separate summations over each index $\mu_1$ and $\mu_2$. To avoid confusion, whenever summed, a multi-index subscript will always be styled with the same Greek letter distinctly numbered. There is no real difference between such a subscript and the spacetime index used to indicate differentiation of fields; styling them differently is merely an indication of what they are summed over, so the styling only matters when indices are repeated.

Let us proceed to construct a non-linear connection on $J^\mathsf{q}(\pi)$. The coefficients in the d-tensor derivatives
\begin{equation}
    \frac{\delta}{\delta \phi^i_{\mu_R}} = \frac{\partial}{\partial \phi^i_{\mu_\mathsf{R}}} - N^{j \mu_\mathsf{R}}_{\rho_{\mathsf{R}+1} i} \frac{\partial}{\partial \phi^j_{\rho_{\mathsf{R}+1}}} - N^{j \mu_\mathsf{R}}_{\rho_{\mathsf{R}+2} i} \frac{\partial}{\partial \phi^j_{\rho_{\mathsf{R}+2}}} - \ldots - N^{j \mu_\mathsf{R}}_{\rho_\mathsf{Q} i} \frac{\partial}{\partial \phi^j_{\rho_\mathsf{Q}}} \, ,
\end{equation}
must satisfy the transformation laws
\begin{equation}
    N^{k\mu_\mathsf{S}}_{\rho_\mathsf{R} l} \frac{\partial \phi^l}{\partial \tilde{\phi}^i} = \tilde{N}^{j\mu_\mathsf{S}}_{\rho_\mathsf{R} i} \frac{\partial \phi^k}{\partial \tilde{\phi}^j} + \tilde{N}^{j\mu_\mathsf{S}}_{\tau_{\mathsf{R}-1} i} \frac{\partial \phi^k_{\rho_\mathsf{R}}}{\partial \tilde{\phi}^j_{\tau_{\mathsf{R}-1}}} + \ldots + \tilde{N}^{j\mu_\mathsf{S}}_{\tau_{\mathsf{S}+1} i} \frac{\partial \phi^k_{\rho_\mathsf{R}}}{\partial \tilde{\phi}^j_{\tau_{\mathsf{S}+1}}} - \frac{\partial \phi^k_{\rho_\mathsf{R}}}{\partial \tilde{\phi}^i_{\mu_\mathsf{S}}} \, .
\end{equation}
In the general case $\mathsf{q} \geq 2$, it is easier to work with the dual basis
\begin{equation}
    \delta \phi^i_{\mu_\mathsf{R}} = d\phi^i_{\mu_\mathsf{R}} + M^{i \rho_{\mathsf{R}-1}}_{\mu_\mathsf{R} j} d\phi^j_{\rho_{\mathsf{R}-1}} + \ldots + M^{i \rho}_{\mu_\mathsf{R} j} d\phi^j_\rho + M^i_{\mu_\mathsf{R} j} d\phi^j \, ,
\end{equation}
whose coefficients $M$ determine $N$ recursively via duality:
\begin{equation}
    M^{i \rho_\mathsf{S}}_{\mu_\mathsf{R} j} = N^{i \rho_\mathsf{S}}_{\mu_\mathsf{R} j} + M^{i \sigma_{\mathsf{R}-1}}_{\mu_\mathsf{R} k} N^{k \rho_\mathsf{S}}_{\sigma_{\mathsf{R}-1} j} + \ldots + M^{i \sigma_{\mathsf{S}+1}}_{\mu_\mathsf{R} k} N^{k \rho_\mathsf{S}}_{\sigma_{\mathsf{S}+1} j} \, ,
\end{equation}
and must transform as
\begin{equation}
    \tilde{M}^{i \rho_\mathsf{S}}_{\mu_\mathsf{R} j} \frac{\partial \phi^k}{\partial \tilde{\phi}^i} = M^{k \rho_\mathsf{S}}_{\mu_\mathsf{R} l} \frac{\partial \phi^l}{\partial \tilde{\phi}^j} + M^{k \tau_{\mathsf{S}+1}}_{\mu_\mathsf{R} l} \frac{\partial \phi^l_{\tau_{\mathsf{S}+1}}}{\partial \tilde{\phi}^j_{\rho_\mathsf{S}}} + \ldots + M^{k \tau_{\mathsf{R}-1}}_{\mu_\mathsf{R} l} \frac{\partial \phi^l_{\tau_{\mathsf{R}-1}}}{\partial \tilde{\phi}^j_{\rho_\mathsf{S}}} + \frac{\partial \phi^k_{\mu_\mathsf{R}}}{\partial \tilde{\phi}^j_{\rho_\mathsf{S}}} \, .
\end{equation}

The operator
\begin{equation}
    \Gamma^{(\mathsf{q})}_{\tau_0} = \frac{\partial}{\partial x^{\tau_0}} + \phi^i_{\tau_0} \frac{\partial}{\partial \phi^i} + \phi^i_{\tau_0 \tau_1} \frac{\partial}{\partial \phi^i_{\tau_1}} + \phi^i_{\tau_0 \tau_1 \tau_2} \frac{\partial}{\partial \phi^i_{\tau_1 \tau_2}} + \ldots + \phi^i_{\tau_0 \tau_{\mathsf{Q}-1}} \frac{\partial}{\partial \phi^i_{\tau_{\mathsf{Q}-1}}} \, ,
\end{equation}
is the truncation of $d/dx^{\tau_0}$ to order $\mathsf{q}$ on $J^\mathsf{q}(\pi)$. It acts as follows:
\begin{equation}
    \tilde{\Gamma}^{(\mathsf{q})}_{\tau_\mathsf{r}} \frac{\partial \phi^k_{\tau_{\mathsf{R}-1}}}{\partial \tilde{\phi}^j_{\rho_\mathsf{S}}} = \frac{\partial}{\partial \tilde{\phi}^j_{\rho_\mathsf{S}}} \left ( \tilde{\Gamma}^{(\mathsf{q})}_{\tau_\mathsf{r}} \phi^k_{\tau_{\mathsf{R}-1}} \right ) - \delta^{\rho_\mathsf{S}}_{\tau_\mathsf{r} \sigma_{\mathsf{S}-1}} \frac{\partial \phi^k_{\tau_{\mathsf{R}-1}}}{\partial \tilde{\phi}^j_{\sigma_{\mathsf{S}-1}}} = \frac{\partial \phi^k_{\tau_\mathsf{R}}}{\partial \tilde{\phi}^j_{\rho_\mathsf{S}}} - \sum_{\mathsf{r} \in \mathsf{I} \subseteq \mathsf{R}} \delta^{\rho_\mathsf{S}}_{\tau_\mathsf{I}} \frac{\partial \phi^k_{\tau_\mathsf{R} - \tau_\mathsf{I}}}{\partial \tilde{\phi}^j} \, ,
\end{equation}
because every partition of $\mathsf{R}$ is generated from $\mathsf{R}-1$ by either appending $\mathsf{r}$ to one subset or creating a new singleton comprising $\mathsf{r}$. Its transformation law yields a $\partial / \partial \phi^k_{\sigma_\mathsf{Q}}$ piece:
\begin{equation}
    \tilde{\Gamma}^{(\mathsf{q})}_{\tau_0} = \Gamma^{(\mathsf{q})}_{\tau_0} + \left (\tilde{\phi}^i_{\tau_0} \frac{\partial \phi^k_{\sigma_\mathsf{Q}}}{\partial \tilde{\phi}^i} + \tilde{\phi}^i_{\tau_0 \tau_1} \frac{\partial \phi^k_{\sigma_\mathsf{Q}}}{\partial \tilde{\phi}^i_{\tau_1}} + \ldots \tilde{\phi}^i_{\tau_0 \tau_{\mathsf{Q}-1}} \frac{\partial \phi^k_{\sigma_\mathsf{Q}}}{\partial \tilde{\phi}^i_{\tau_{\mathsf{Q}-1}}} \right ) \frac{\partial}{\partial \phi^k_{\sigma_\mathsf{Q}}} \, .
\end{equation}
This can be obtained by computing
\begin{equation}
\begin{aligned}
    \tilde{\Gamma}^{(\mathsf{q})}_{\tau_0} - \Gamma^{(\mathsf{q})}_{\tau_0} &= \sum_{\mathsf{s}=0}^{\mathsf{q}-1} \left [ \left ( \frac{\partial \tilde{\phi}^i_{\sigma_\mathsf{S}}}{\partial \phi^j_{\tau_{\mathsf{S}-1}}} \phi^j_{\tau_0 \tau_{\mathsf{S}-1}} + \ldots + \frac{\partial \tilde{\phi}^i_{\sigma_\mathsf{S}}}{\partial \phi^j} \phi^j_{\tau_0} \right ) \frac{\partial \phi^k}{\partial \tilde{\phi}^i} \frac{\partial}{\partial \phi^k_{\sigma_\mathsf{S}}} + \tilde{\phi}^i_{\tau_0 \tau_\mathsf{S}} \left ( \frac{\partial \phi^k_{\sigma_{\mathsf{S}+1}}}{\partial \tilde{\phi}^i_{\tau_\mathsf{S}}} \frac{\partial}{\partial \phi^k_{\sigma_{\mathsf{S}+1}}} + \ldots + \frac{\partial \phi^k_{\sigma_\mathsf{Q}}}{\partial \tilde{\phi}^i_{\tau_\mathsf{S}}} \frac{\partial}{\partial \phi^k_{\sigma_\mathsf{Q}}} \right ) \right ] \\
    &= \sum_{\mathsf{r}=0}^{\mathsf{q}-1} \left [ \left ( \frac{\partial \tilde{\phi}^i_{\sigma_\mathsf{R}}}{\partial \phi^j_{\tau_{\mathsf{R}-1}}} \phi^j_{\tau_0 \tau_{\mathsf{R}-1}} + \ldots + \frac{\partial \tilde{\phi}^i_{\sigma_\mathsf{R}}}{\partial \phi^j} \phi^j_{\tau_0} \right ) \frac{\partial \phi^k}{\partial \tilde{\phi}^i} + \left ( \tilde{\phi}^i_{\tau_0 \tau_{\mathsf{R}-1}} \frac{\partial \phi^k_{\sigma_\mathsf{R}}}{\partial \tilde{\phi}^i_{\tau_{\mathsf{R}-1}}} + \ldots + \tilde{\phi}^i_{\tau_0} \frac{\partial \phi^k_{\sigma_\mathsf{R}}}{\partial \tilde{\phi}^i} \right ) \right ] \frac{\partial}{\partial \phi^k_{\sigma_\mathsf{R}}} \\
    &\quad + \left [ (0) + \left ( \tilde{\phi}^i_{\tau_0 \tau_{\mathsf{Q}-1}} \frac{\partial \phi^k_{\sigma_\mathsf{Q}}}{\partial \tilde{\phi}^i_{\tau_{\mathsf{Q}-1}}} + \ldots + \tilde{\phi}^i_{\tau_0} \frac{\partial \phi^k_{\sigma_\mathsf{Q}}}{\partial \tilde{\phi}^i} \right ) \right ] \frac{\partial}{\partial \phi^k_{\sigma_\mathsf{Q}}} \, .
\end{aligned}
\end{equation}
In the sum over $\mathsf{r}$, we collected terms according to their derivatives $\partial / \partial \phi^k_{\sigma_\mathsf{R}}$, with $\mathsf{r} = \mathsf{q}$ missing the first part. One can explicitly check that the summand vanishes for $\mathsf{r} < \mathsf{q}$. Alternatively, on the infinite jet bundle $\mathsf{q} \rightarrow \infty$, we must have $\lim \Gamma^{(\mathsf{q})}_{\tau_0} \equiv \Gamma_{\tau_0} = d/dx^{\tau_0} = \tilde{\Gamma}_{\tau_0}$, which means that all the summands for $\mathsf{r} < \mathsf{q}$ vanish since each is independent of $\mathsf{q}$. Therefore, it holds on $J^\mathsf{q}(\pi)$ that given any object $T(x^\alpha, \phi^a, \phi^a_\Lambda)$ that is independent of $\phi^a_\Lambda$ for $|\Lambda| = \mathsf{q}$, we have the simple transformation law
\begin{equation}
    \Gamma^{(\mathsf{q})}_{\tau_0} T = \tilde{\Gamma}^{(\mathsf{q})}_{\tau_0} T \, .
\end{equation}

We construct a recursive solution for the $M$'s. The edge case $\mathsf{s} = \mathsf{r}-1$ is simple. The transformation law
\begin{equation}
    \tilde{M}^{i \rho_{\mathsf{R}-1}}_{\mu_\mathsf{R} j} \frac{\partial \phi^k}{\partial \tilde{\phi}^i} = M^{k \rho_{\mathsf{R}-1}}_{\mu_\mathsf{R} l} \frac{\partial \phi^l}{\partial \tilde{\phi}^j} + \frac{\partial \phi^k_{\mu_\mathsf{R}}}{\partial \tilde{\phi}^j_{\rho_{\mathsf{R}-1}}} = M^{k \rho_{\mathsf{R}-1}}_{\mu_\mathsf{R} l} \frac{\partial \phi^l}{\partial \tilde{\phi}^j} + \sum_{\mathsf{a} \in \mathsf{R}} \delta^{\rho_{\mathsf{R}-1}}_{\mu_\mathsf{R} - \mu_\mathsf{a}} \frac{\partial \phi^k_{\mu_\mathsf{a}}}{\partial \tilde{\phi}^j} \, ,
\end{equation}
is satisfied by
\begin{equation}
    M^{i \rho_{\mathsf{R}-1}}_{\mu_\mathsf{R} j} = \sum_{\mathsf{a} \in \mathsf{R}} M^i_{\mu_\mathsf{a} j} \delta^{\rho_{\mathsf{R}-1}}_{\mu_\mathsf{R} - \mu_\mathsf{a}} \, .
\end{equation}
Meanwhile, our guess for $\mathsf{s} < \mathsf{r}-1$ will be
\begin{equation}
\begin{aligned}
    M^{i \rho_\mathsf{S}}_{\mu_\mathsf{R} j} &= \frac{\mathsf{r}}{\mathsf{r}-\mathsf{s}} \left [ \Gamma^{(\mathsf{q})}_{(\mu_\mathsf{r}} M^{i \rho_\mathsf{S}}_{\mu_{\mathsf{R}-1}) j} + M^i_{(\mu_\mathsf{r}| m} M^{m \rho_\mathsf{S}}_{|\mu_{\mathsf{R}-1}) j} \right ] \\
    &= \binom{\mathsf{r}}{\mathsf{s}} \sum_{\mathcal{O} \text{'s} \, \in \{ \Gamma, M \}} \mathcal{O}^i_{(\mu_{\mathsf{s}+1} | m_{\mathsf{s}+1} } \ldots \mathcal{O}^{m_{\mathsf{r}-2}}_{|\mu_{\mathsf{r}-1}| m_{\mathsf{r}-1}} \mathcal{O}^{m_{\mathsf{r}-1}}_{|\mu_\mathsf{r}|j} \delta^{\rho_\mathsf{S}}_{|\mu_\mathsf{S})} \, .
\end{aligned}
\end{equation}
This is to be verified by induction, descending on $\mathsf{s}$ before ascending on $\mathsf{r}$ up to $\mathsf{q}$. The second line results from applying the first line recursively until $\mathsf{r}$ descends to $\mathsf{s}+1$. By the (normalized) symmetrization of the $\mu$'s, we refer to that which yields $\mathsf{r}!$ terms. Each operation $\mathcal{O}^m_{\mu n}$ can be either differentiation by $\delta^m_n \Gamma_\mu$ or contraction with $M^m_{\mu n}$, the former acting on everything to the right. Note that according to our guess, $M^{i \rho_\mathsf{S}}_{\mu_{\mathsf{R}-1} j}$ is independent of $\phi^\alpha_\Lambda$ for $|\Lambda| \geq \mathsf{r}-\mathsf{s}$, so $\Gamma_{\mu_\mathsf{r}}$ transforms simply when acting on any that appears in the induction.

Suppose the inductive hypothesis holds up to $\mathsf{s}+1$. Applying $\tilde{\Gamma}_{\mu_\mathsf{r}}$ to the transformation law of $\tilde{M}^{i \rho_\mathsf{S}}_{\mu_{\mathsf{R}-1} j}$, we get
\begin{equation} \label{eq:GammaM}
\begin{aligned}
    \left ( \tilde{\Gamma}_{\mu_\mathsf{r}} \tilde{M}^{i \rho_\mathsf{S}}_{\mu_{\mathsf{R}-1} j} \right ) \frac{\partial \phi^k}{\partial \tilde{\phi}^i} + \tilde{M}^{i \rho_\mathsf{S}}_{\mu_{\mathsf{R}-1} j} \frac{\partial \phi^k_{\mu_\mathsf{r}}}{\partial \tilde{\phi}^i} &= \left ( \Gamma_{\mu_\mathsf{r}} M^{k \rho_\mathsf{S}}_{\mu_{\mathsf{R}-1} l} \right ) \frac{\partial \phi^l}{\partial \tilde{\phi}^j} + \left ( \Gamma_{\mu_\mathsf{r}} M^{k \tau_{\mathsf{S}+1}}_{\mu_{\mathsf{R}-1} l} \right ) \frac{\partial \phi^l_{\tau_{\mathsf{S}+1}}}{\partial \tilde{\phi}^j_{\rho_\mathsf{S}}} + \ldots + \left ( \Gamma_{\mu_\mathsf{r}} M^{k \tau_{\mathsf{R}-2}}_{\mu_{\mathsf{R}-1} l} \right ) \frac{\partial \phi^l_{\tau_{\mathsf{R}-2}}}{\partial \tilde{\phi}^j_{\rho_\mathsf{S}}} \\
    &\quad + \frac{\partial \phi^k_{\mu_\mathsf{R}}}{\partial \tilde{\phi}^j_{\rho_\mathsf{S}}} + M^{k \rho_\mathsf{S}}_{\mu_{\mathsf{R}-1} l} \frac{\partial \phi^l_{\mu_\mathsf{r}}}{\partial \tilde{\phi}^j} + M^{k \tau_{\mathsf{S}+1}}_{\mu_{\mathsf{R}-1} l} \left ( \frac{\partial \phi^l_{\tau_{\mathsf{S}+1} \mu_\mathsf{r}}}{\partial \tilde{\phi}^j_{\rho_\mathsf{S}}} - \sum_{\mathsf{I} \subseteq \mathsf{S}+1} \delta^{\rho_\mathsf{S}}_{\tau_\mathsf{I} \mu_\mathsf{r}} \frac{\partial \phi^l_{\tau_{\mathsf{S}+1} - \tau_\mathsf{I}}}{\partial \tilde{\phi}^j} \right ) + \\
    &\quad \ldots + M^{k \tau_{\mathsf{R}-2}}_{\mu_{\mathsf{R}-1} l} \left ( \frac{\partial \phi^l_{\tau_{\mathsf{R}-2} \mu_\mathsf{r}}}{\partial \tilde{\phi}^j_{\rho_\mathsf{S}}} - \sum_{\mathsf{I} \subseteq \mathsf{R}-2} \delta^{\rho_\mathsf{S}}_{\tau_\mathsf{I} \mu_\mathsf{r}} \frac{\partial \phi^l_{\tau_{\mathsf{R}-2} - \tau_I}}{\partial \tilde{\phi}^j} \right ) - \sum_{\mathsf{r} \in \mathsf{I} \subseteq \mathsf{R}} \delta^{\rho_\mathsf{S}}_{\mu_\mathsf{I}} \frac{\partial \phi^k_{\mu_\mathsf{R} - \mu_\mathsf{I}}}{\partial \tilde{\phi}^j} \, .
\end{aligned}
\end{equation}
Meanwhile, contracting the transformation laws of $\tilde{M}^i_{\mu_\mathsf{r} m}$ and $\tilde{M}^{m \rho_\mathsf{S}}_{\mu_{\mathsf{R}-1} j}$, we get
\begin{equation} \label{eq:MM}
\begin{aligned}
    \tilde{M}^i_{\mu_\mathsf{r} m} \tilde{M}^{m \rho_\mathsf{S}}_{\mu_{\mathsf{R}-1} j} \frac{\partial \phi^k}{\partial \tilde{\phi}^i} &= \left ( \tilde{M}^i_{\mu_\mathsf{r} m} \frac{\partial \phi^k}{\partial \tilde{\phi}^i} \right ) \frac{\partial \tilde{\phi}^m}{\partial \phi^n} \left ( \tilde{M}^{p \rho_\mathsf{S}}_{\mu_{\mathsf{R}-1} j} \frac{\partial \phi^n}{\partial \tilde{\phi}^p} \right ) \\
    &= M^k_{\mu_\mathsf{r} m} M^{m \rho_\mathsf{S}}_{\mu_{\mathsf{R}-1} l} \frac{\partial \phi^l}{\partial \tilde{\phi}^j} + M^k_{\mu_\mathsf{r} m} M^{m \tau_{\mathsf{S}+1}}_{\mu_{\mathsf{R}-1} l} \frac{\partial \phi^l_{\tau_{\mathsf{S}+1}}}{\partial \tilde{\phi}^j_{\rho_\mathsf{S}}} + \ldots + M^k_{\mu_\mathsf{r} m} M^{m \tau_{\mathsf{R}-2}}_{\mu_{\mathsf{R}-1} l} \frac{\partial \phi^l_{\tau_{\mathsf{R}-2}}}{\partial \tilde{\phi}^j_{\rho_\mathsf{S}}} + M^k_{\mu_\mathsf{r} l} \frac{\partial \phi^l_{\mu_{\mathsf{R}-1}}}{\partial \tilde{\phi}^j_{\rho_\mathsf{S}}} \\
    &\quad + \left ( M^{n \rho_\mathsf{S}}_{\mu_{\mathsf{R}-1} q} \frac{\partial \phi^q}{\partial \tilde{\phi}^j} + M^{n \tau_{\mathsf{S}+1}}_{\mu_{\mathsf{R}-1} q} \frac{\partial \phi^q_{\tau_{\mathsf{S}+1}}}{\partial \tilde{\phi}^j_{\rho_\mathsf{S}}} + \ldots + M^{n \tau_{\mathsf{R}-2}}_{\mu_{\mathsf{R}-1} q} \frac{\partial \phi^q_{\tau_{\mathsf{R}-2}}}{\partial \tilde{\phi}^j_{\rho_\mathsf{S}}} + \frac{\partial \phi^n_{\mu_{\mathsf{R}-1}}}{\partial \tilde{\phi}^j_{\rho_\mathsf{S}}} \right ) \frac{\partial \tilde{\phi}^i}{\partial \phi^n} \frac{\partial \phi^k_{\mu_\mathsf{r}}}{\partial \tilde{\phi}^i} \, .
\end{aligned}
\end{equation}
We also have
\begin{equation} \label{eq:Mx}
    \tilde{M}^{i \rho_\mathsf{S}}_{\mu_{\mathsf{R}-1} j} \frac{\partial \phi^k_{\mu_\mathsf{r}}}{\partial \tilde{\phi}^i} = \left ( M^{n \rho_\mathsf{S}}_{\mu_{\mathsf{R}-1} q} \frac{\partial \phi^q}{\partial \tilde{\phi}^j} + M^{n \tau_{\mathsf{S}+1}}_{\mu_{\mathsf{R}-1} q} \frac{\partial \phi^q_{\tau_{\mathsf{S}+1}}}{\partial \tilde{\phi}^j_{\rho_\mathsf{S}}} + \ldots + M^{n \tau_{\mathsf{R}-2}}_{\mu_{\mathsf{R}-1} q} \frac{\partial \phi^q_{\tau_{\mathsf{R}-2}}}{\partial \tilde{\phi}^j_{\rho_\mathsf{S}}} + \frac{\partial \phi^n_{\mu_{\mathsf{R}-1}}}{\partial \tilde{\phi}^j_{\rho_\mathsf{S}}} \right ) \frac{\partial \tilde{\phi}^i}{\partial \phi^n} \frac{\partial \phi^k_{\mu_\mathsf{r}}}{\partial \tilde{\phi}^i} \, .
\end{equation}
Now symmetrize all $\mu$'s and subtract \cref{eq:Mx} from the sum of \cref{eq:GammaM} and \cref{eq:MM} to obtain
\begin{equation} \label{eq:extraneous}
\begin{aligned}
    &\frac{\mathsf{r}-\mathsf{s}}{\mathsf{r}} \tilde{M}^{i \rho_\mathsf{S}}_{\mu_\mathsf{R} j} \frac{\partial \phi^k}{\partial \tilde{\phi}^i} = \frac{\mathsf{r}-\mathsf{s}}{\mathsf{r}} M^{k \rho_\mathsf{S}}_{\mu_\mathsf{R} l} \frac{\partial \phi^l}{\partial \tilde{\phi}^j} + \frac{\mathsf{r}-\mathsf{s}-1}{\mathsf{r}} M^{k \tau_{\mathsf{S}+1}}_{\mu_\mathsf{R} l} \frac{\partial \phi^l_{\tau_{\mathsf{S}+1}}}{\partial \tilde{\phi}^j_{\rho_\mathsf{S}}} + \ldots + \frac{2}{\mathsf{r}} M^{k \tau_{\mathsf{R}-2}}_{\mu_\mathsf{R} l} \frac{\partial \phi^l_{\tau_{\mathsf{R}-2}}}{\partial \tilde{\phi}^j_{\rho_\mathsf{S}}} \\
    &\qquad + M^{k \rho_\mathsf{S}}_{(\mu_{\mathsf{R}-1}| l} \frac{\partial \phi^l_{|\mu_\mathsf{r})}}{\partial \tilde{\phi}^j} + M^{k \tau_{\mathsf{S}+1}}_{(\mu_{\mathsf{R}-1}| l} \left ( \frac{\partial \phi^l_{\tau_{\mathsf{S}+1} |\mu_\mathsf{r})}}{\partial \tilde{\phi}^j_{\rho_\mathsf{S}}} - \sum_{\mathsf{I} \subseteq \mathsf{S}+1} \delta^{\rho_\mathsf{S}}_{\tau_\mathsf{I} |\mu_\mathsf{r})} \frac{\partial \phi^l_{\tau_{\mathsf{S}+1} - \tau_\mathsf{I}}}{\partial \tilde{\phi}^j} \right ) + \ldots \\
    &\qquad + M^{k \tau_{\mathsf{R}-2}}_{(\mu_{\mathsf{R}-1}| l} \left ( \frac{\partial \phi^l_{\tau_{\mathsf{R}-2} |\mu_\mathsf{r})}}{\partial \tilde{\phi}^j_{\rho_\mathsf{S}}} - \sum_{\mathsf{I} \subseteq \mathsf{R}-2} \delta^{\rho_\mathsf{S}}_{\tau_\mathsf{I} |\mu_\mathsf{r})} \frac{\partial \phi^l_{\tau_{\mathsf{R}-2} - \tau_\mathsf{I}}}{\partial \tilde{\phi}^j} \right ) + M^k_{(\mu_\mathsf{r}| l} \frac{\partial \phi^l_{|\mu_{\mathsf{R}-1})}}{\partial \tilde{\phi}^j_{\rho_\mathsf{S}}} + \left ( \frac{\partial \phi^k_{\mu_\mathsf{R}}}{\partial \tilde{\phi}^j_{\rho_\mathsf{S}}} - \sum_{\mathsf{r} \in \mathsf{I} \subseteq \mathsf{R}} \delta^{\rho_\mathsf{S}}_{\mu_\mathsf{I}} \frac{\partial \phi^k_{\mu_\mathsf{R} - \mu_\mathsf{I}}}{\partial \tilde{\phi}^j} \right ) \, .
\end{aligned}
\end{equation}

Let us tackle the extra terms on the second and third lines of \cref{eq:extraneous}. Out of a total of $\binom{\mathsf{r}}{\mathsf{s}}$ partitions of $\mu_\mathsf{R}$ into $\mu_\mathsf{I}$ and $\mu_\mathsf{R} - \mu_\mathsf{I}$ with $|\mu_\mathsf{I}| = \mathsf{s}$, there are $\binom{\mathsf{r}-1}{\mathsf{s}-1} = \frac{\mathsf{s}}{\mathsf{r}} \binom{\mathsf{r}}{\mathsf{s}}$ with $\mathsf{r} \in \mathsf{I}$. Hence,
\begin{equation}
    \frac{\partial \phi^k_{\mu_\mathsf{R}}}{\partial \tilde{\phi}^j_{\rho_\mathsf{S}}} - \sum_{\mathsf{r} \in \mathsf{I} \subseteq \mathsf{R}} \delta^{\rho_\mathsf{S}}_{(\mu_\mathsf{I}} \frac{\partial \phi^k_{\mu_\mathsf{R} - \mu_\mathsf{I})}}{\partial \tilde{\phi}^j} = \frac{\mathsf{r}-\mathsf{s}}{\mathsf{r}} \frac{\partial \phi^k_{\mu_\mathsf{R}}}{\partial \tilde{\phi}^j_{\rho_\mathsf{S}}} \, .
\end{equation}
Moreover,
\begin{equation}
\begin{aligned}
    M^k_{(\mu_\mathsf{r}| l} \frac{\partial \phi^l_{|\mu_{\mathsf{R}-1})}}{\partial \tilde{\phi}^j_{\rho_\mathsf{S}}} &= \sum_{\mathsf{I} \subseteq \mathsf{R}-1} M^k_{(\mu_\mathsf{r}| l} \delta^{\rho_\mathsf{S}}_{|\mu_\mathsf{I}} \frac{\partial \phi^l_{\mu_{\mathsf{R}-1} - \mu_\mathsf{I})}}{\partial \tilde{\phi}^j} \\
    &= \frac{1}{\mathsf{r}} \sum_{\mathsf{a} \in \mathsf{R}} \sum_{\mathsf{I} \subseteq \mathsf{R}-1} M^k_{(\mu_\mathsf{a}| l} \delta^{\tau_{\mathsf{R}-1}}_{|\mu_\mathsf{R} - \mu_\mathsf{a})} \delta^{\rho_\mathsf{S}}_{\tau_\mathsf{I}} \frac{\partial \phi^i_{\tau_{\mathsf{R}-1} - \tau_\mathsf{I}}}{\partial \tilde{\phi}^j} = \frac{1}{r} M^{k \tau_{\mathsf{R}-1}}_{\mu_\mathsf{R} l} \frac{\partial \phi^l_{\tau_{\mathsf{R}-1}}}{\partial \tilde{\phi}^j_{\rho_\mathsf{S}}} \, .
\end{aligned}
\end{equation}
Writing $\mathsf{S} + \mathsf{N}$ for $\{1, \ldots, \mathsf{s}+\mathsf{n} \}$ where $0 \leq \mathsf{n} \leq \mathsf{r}-\mathsf{s}-2$, we also know from the hypothesis that
\begin{equation}
\begin{aligned}
    &\quad \, M^{k \tau_{\mathsf{S}+\mathsf{N}}}_{(\mu_{\mathsf{R}-1}| l} \left ( \frac{\partial \phi^l_{\tau_{\mathsf{S}+\mathsf{N}} |\mu_\mathsf{r})}}{\partial \tilde{\phi}^j_{\rho_\mathsf{S}}} - \sum_{\mathsf{I} \subseteq \mathsf{S}+\mathsf{N}} \delta^{\rho_\mathsf{S}}_{\tau_\mathsf{I} |\mu_\mathsf{r})} \frac{\partial \phi^l_{\tau_{\mathsf{S}+\mathsf{N}} - \tau_\mathsf{I}}}{\partial \tilde{\phi}^j} \right ) \\
    &= \binom{\mathsf{r}-1}{\mathsf{s}+\mathsf{n}} \sum_{\mathcal{O} \mathrm{'s} \, \in \{ \Gamma, M \}} \mathcal{O}^k_{(\mu_{\mathsf{s}+\mathsf{n}+1} | m_{\mathsf{s}+\mathsf{n}+1} } \ldots \mathcal{O}^{m_{\mathsf{r}-3}}_{|\mu_{\mathsf{r}-2}| m_{\mathsf{r}-2}} \mathcal{O}^{m_{\mathsf{r}-2}}_{|\mu_{\mathsf{r}-1}|l} \sum_{\mathsf{I} \subseteq \mathsf{S}+\mathsf{N}} \delta^{\tau_{\mathsf{S}+\mathsf{N}}}_{|\mu_{\mathsf{S}+\mathsf{N}}|} \delta^{\rho_\mathsf{S}}_{\tau_\mathsf{I}} \frac{\partial \phi^l_{\tau_{\mathsf{S}+\mathsf{N}} - \tau_\mathsf{I} + |\mu_\mathsf{r})}}{\partial \tilde{\phi}^j} \\
    &= \frac{\mathsf{s}+\mathsf{n}+1}{\mathsf{r}} \binom{\mathsf{r}}{\mathsf{s}+\mathsf{n}+1} \times \frac{\mathsf{n}+1}{\mathsf{s}+\mathsf{n}+1} \times \\
    &\hspace{49.3pt} \sum_{\mathcal{O} \mathrm{'s} \, \in \{ \Gamma, M \}} \mathcal{O}^k_{(\mu_{\mathsf{s}+\mathsf{n}+1} | m_{\mathsf{s}+\mathsf{n}+1} } \ldots \mathcal{O}^{m_{\mathsf{r}-3}}_{|\mu_{\mathsf{r}-2}| m_{\mathsf{r}-2}} \mathcal{O}^{m_{\mathsf{r}-2}}_{|\mu_{\mathsf{r}-1}|l} \sum_{\mathsf{J} \subseteq \mathsf{S}+\mathsf{N}+1} \delta^{\tau_{\mathsf{S}+\mathsf{N}+1}}_{|\mu_{\mathsf{S}+\mathsf{N}} \mu_\mathsf{r})} \delta^{\rho_\mathsf{S}}_{\tau_\mathsf{J}} \frac{\partial \phi^l_{\tau_{\mathsf{S}+\mathsf{N}+1} - \tau_\mathsf{J}}}{\partial \tilde{\phi}^j} \\
    &= \frac{\mathsf{n}+1}{\mathsf{r}} M^{k \tau_{\mathsf{S}+\mathsf{N}+1}}_{\mu_\mathsf{R} l} \frac{\partial \phi^l_{\tau_{\mathsf{S}+\mathsf{N}+1}}}{\partial \tilde{\phi}^j_{\rho_\mathsf{S}}} \, ,
\end{aligned}
\end{equation}
because there are $\binom{\mathsf{s}+\mathsf{n}+1}{\mathsf{s}}$ choices of $\mathsf{J} \subseteq \mathsf{S}+\mathsf{N}+1$ with $|\mathsf{J}| = \mathsf{s}$, as opposed to $\binom{\mathsf{s}+\mathsf{n}}{\mathsf{s}} = \frac{\mathsf{n}+1}{\mathsf{s}+\mathsf{n}+1} \binom{\mathsf{s}+\mathsf{n}+1}{\mathsf{s}}$ choices of $\mathsf{I} \subseteq \mathsf{S}+\mathsf{N}$ with $|\mathsf{I}| = \mathsf{s}$. Note that the special case $\mathsf{n} = 0$ yields
\begin{equation}
    M^{k \rho_\mathsf{S}}_{(\mu_{\mathsf{R}-1}| l} \frac{\partial \phi^l_{|\mu_\mathsf{r})}}{\partial \tilde{\phi}^j} = M^{k \tau_\mathsf{S}}_{(\mu_{\mathsf{R}-1}| l} \left ( \frac{\partial \phi^l_{\tau_\mathsf{S} |\mu_\mathsf{r})}}{\partial \tilde{\phi}^j_{\rho_\mathsf{S}}} - \sum_{\mathsf{I} \subseteq \mathsf{S}} \delta^{\rho_\mathsf{S}}_{\tau_\mathsf{I} |\mu_\mathsf{r})} \frac{\partial \phi^l_{\tau_\mathsf{S} - \tau_\mathsf{I}}}{\partial \tilde{\phi}^j} \right ) = \frac{1}{r} M^{k \tau_{\mathsf{S}+1}}_{\mu_\mathsf{R} l} \frac{\partial \phi^l_{\tau_{\mathsf{S}+1}}}{\partial \tilde{\phi}^j_{\rho_\mathsf{S}}} \, .
\end{equation}
Therefore, the extra terms in \cref{eq:extraneous} can be combined with its first line to give
\begin{equation}
    \frac{\mathsf{r}-\mathsf{s}}{\mathsf{r}} \tilde{M}^{i \rho_\mathsf{S}}_{\mu_\mathsf{R} j} \frac{\partial \phi^k}{\partial \tilde{\phi}^i} = \frac{\mathsf{r}-\mathsf{s}}{\mathsf{r}} \left ( M^{k \rho_\mathsf{S}}_{\mu_\mathsf{R} l} \frac{\partial \phi^l}{\partial \tilde{\phi}^j} + M^{k \tau_{\mathsf{S}+1}}_{\mu_\mathsf{R} l} \frac{\partial \phi^l_{\tau_{\mathsf{S}+1}}}{\partial \tilde{\phi}^j_{\rho_\mathsf{S}}} + \ldots + M^{k \tau_{\mathsf{R}-1}}_{\mu_\mathsf{R} l} \frac{\partial \phi^l_{\tau_{\mathsf{R}-1}}}{\partial \tilde{\phi}^j_{\rho_\mathsf{S}}} + \frac{\partial \phi^k_{\mu_\mathsf{R}}}{\partial \tilde{\phi}^j_{\rho_\mathsf{S}}} \right ) \, ,
\end{equation}
which is the desired transformation law of $M^{i \rho_\mathsf{S}}_{\mu_\mathsf{R} j}$ and hence completes the induction.

We can now explicitly write down a non-linear connection on jet bundles of arbitrary order. Starting with
\begin{equation}
    M^i_{\mu j} = N^i_{\mu j} = \gamma^i_{\;jk} \phi^k_\mu, \quad M^{i \rho_{\mathsf{R}-1}}_{\mu_\mathsf{R} j} = \sum_{\mathsf{a} \in \mathsf{R}} M^i_{\mu_\mathsf{a} j} \delta^{\rho_{\mathsf{R}-1}}_{\mu_\mathsf{R} - \mu_\mathsf{a}} \, ,
\end{equation}
and recalling the recursive relations
\begin{equation}
\begin{aligned}
    M^{i \rho_\mathsf{S}}_{\mu_\mathsf{R} j} &= \frac{\mathsf{r}}{\mathsf{r}-\mathsf{s}} \left [ \Gamma_{(\mu_\mathsf{r}} M^{i \rho_\mathsf{S}}_{\mu_{\mathsf{R}-1}) j} + M^i_{(\mu_\mathsf{r}| m} M^{m \rho_\mathsf{S}}_{|\mu_{\mathsf{R}-1}) j} \right ] \, , \\
    N^{i \rho_\mathsf{S}}_{\mu_\mathsf{R} j} &= M^{i \rho_\mathsf{S}}_{\mu_\mathsf{R} j} - M^{i \sigma_{\mathsf{R}-1}}_{\mu_\mathsf{R} k} N^{k \rho_\mathsf{S}}_{\sigma_{\mathsf{R}-1} j} - \ldots - M^{i \sigma_{\mathsf{S}+1}}_{\mu_\mathsf{R} k} N^{k \rho_\mathsf{S}}_{\sigma_{\mathsf{S}+1} j} \, ,
\end{aligned}
\end{equation}
we obtain the d-tensor derivatives
\begin{equation}
\begin{alignedat}{10}
    & \frac{\delta}{\delta \phi^i} && = \frac{\partial}{\partial \phi^i} - N^j_{\rho i} && \frac{\partial}{\partial \phi^j_\rho} && - N^j_{\rho_1 \rho_2 i} && \frac{\partial}{\partial \phi^j_{\rho_1 \rho_2}} && - N^j_{\rho_1 \rho_2 \rho_3 i} && \frac{\partial}{\partial \phi^j_{\rho_1 \rho_2 \rho_3}} && - N^j_{\rho_1 \rho_2 \rho_3 \rho_4 i} && \frac{\partial}{\partial \phi^j_{\rho_1 \rho_2 \rho_3 \rho_4}} && - \, \ldots \\
    & \frac{\delta}{\delta \phi^i_\mu} && = && \frac{\partial}{\partial \phi^i_\mu} && - N^{j \mu}_{\rho_1 \rho_2 i} && \frac{\partial}{\partial \phi^j_{\rho_1 \rho_2}} && - N^{j \mu}_{\rho_1 \rho_2 \rho_3 i} && \frac{\partial}{\partial \phi^j_{\rho_1 \rho_2 \rho_3}} && - N^{j \mu}_{\rho_1 \rho_2 \rho_3 \rho_4 i} && \frac{\partial}{\partial \phi^j_{\rho_1 \rho_2 \rho_3 \rho_4}} && - \, \ldots \\
    & \frac{\delta}{\delta \phi^i_{\mu_1 \mu_2}} && = && && && \frac{\partial}{\partial \phi^i_{\mu_1 \mu_2}} && - N^{j \mu_1 \mu_2}_{\rho_1 \rho_2 \rho_3 i} && \frac{\partial}{\partial \phi^j_{\rho_1 \rho_2 \rho_3}} && - N^{j \mu_1 \mu_2}_{\rho_1 \rho_2 \rho_3 \rho_4 i} && \frac{\partial}{\partial \phi^j_{\rho_1 \rho_2 \rho_3 \rho_4}} && - \, \ldots \\
    & \frac{\delta}{\delta \phi^i_{\mu_1 \mu_2 \mu_3}} && = && && && && && \frac{\partial}{\partial \phi^i_{\mu_1 \mu_2 \mu_3}} && - N^{j \mu_1 \mu_2 \mu_3}_{\rho_1 \rho_2 \rho_3 \rho_4 i} && \frac{\partial}{\partial \phi^j_{\rho_1 \rho_2 \rho_3 \rho_4}} && - \, \ldots \\
    & \frac{\delta}{\delta \phi^i_{\mu_1 \mu_2 \mu_3 \mu_4}} && = && && && && && && && \frac{\partial}{\partial \phi^i_{\mu_1 \mu_2 \mu_3 \mu_4}} && - \, \ldots \\
    & && && && && && \vdots
\end{alignedat}
\end{equation}
\interdisplaylinepenalty=10000
where
\begin{alignat*}{2}
    &N^j_{\rho i} &&= \gamma^j_{\;ik} \phi^k_\rho \, , \\
    &N^j_{\rho_1 \rho_2 i} &&= \left ( \gamma^j_{\;ik,l} - \gamma^j_{\;mk} \gamma^m_{\;il} \right ) \phi^k_{(\rho_1} \phi^l_{\rho_2)} + \gamma^j_{\;ik} \phi^k_{\rho_1 \rho_2} \, , \\
    &N^j_{\rho_1 \rho_2 \rho_3 i} &&= \left ( \gamma^j_{\;ik,ln} - 2 \gamma^j_{\;mk,l} \gamma^m_{\;in} - \gamma^j_{\;mk} \gamma^m_{\;il,n} + \gamma^j_{\;mk} \gamma^m_{\;pl} \gamma^p_{\;in} \right ) \phi^k_{(\rho_1} \phi^l_{\rho_2} \phi^n_{\rho_3)} \\
    & &&\quad + \left ( 2 \gamma^j_{\;ik,l} + \gamma^j_{\;il,k} - 2 \gamma^j_{\;mk} \gamma^m_{\;il} - \gamma^j_{\;ml} \gamma^m_{\;ik} \right ) \phi^k_{(\rho_1 \rho_2} \phi^l_{\rho_3 )} + \gamma^j_{\;ik} \phi^k_{\rho_1 \rho_2 \rho_3} \, , \\
    &N^j_{\rho_1 \rho_2 \rho_3 \rho_4 i} &&= \Big ( \gamma^j_{\;ik,lnq} - 3 \gamma^j_{\;mk,ln} \gamma^m_{\;iq} - 3 \gamma^j_{\;mk,l} \gamma^m_{\;in,q} - \gamma^j_{\;mk} \gamma^m_{\;il,nq} + 3 \gamma^j_{\;mk,l} \gamma^m_{\;pn} \gamma^p_{\;iq} + 2 \gamma^j_{\;mk} \gamma^m_{\;pl,n} \gamma^p_{\;iq} + \gamma^j_{\;mk} \gamma^m_{\;pl} \gamma^p_{\;in,q} \\
    & &&\quad - \gamma^j_{\;mk} \gamma^m_{\;pl} \gamma^p_{\;rn} \gamma^r_{\;iq} \Big ) \, \phi^k_{(\rho_1} \phi^l_{\rho_2} \phi^n_{\rho_3} \phi^q_{\rho_4)} + \Big ( 3 \gamma^j_{\;ik,ln} + 3 \gamma^j_{\;il,kn} - 6 \gamma^j_{\;mk,l} \gamma^m_{\;in} - 3 \gamma^j_{\;ml,k} \gamma^m_{\;in} - 3 \gamma^j_{\;ml,n} \gamma^m_{\;ik} \\
    & &&\quad - 3 \gamma^j_{\;mk} \gamma^m_{\;il,n} - 2 \gamma^j_{\;ml} \gamma^m_{\;ik,n} - \gamma^j_{\;ml} \gamma^m_{\;in,k} + 3 \gamma^j_{\;mk} \gamma^m_{\;pl} \gamma^p_{\;in} + 2 \gamma^j_{\;ml} \gamma^m_{\;pk} \gamma^p_{\;in} + \gamma^j_{\;ml} \gamma^m_{\;pn} \gamma^p_{\;ik} \Big ) \, \phi^k_{(\rho_1 \rho_2} \phi^l_{\rho_3} \phi^n_{\rho_4)} \\
    & &&\quad + 3 \left ( \gamma^j_{\;ik,l} - \gamma^j_{\;mk} \gamma^m_{\;il} \right ) \phi^k_{(\rho_1 \rho_2} \phi^l_{\rho_3 \rho_4 )} + \left ( 3 \gamma^j_{\;ik,l} + \gamma^j_{\;il,k} - 3 \gamma^j_{\;mk} \gamma^m_{\;il} - \gamma^j_{\;ml} \gamma^m_{\;ik} \right ) \phi^k_{(\rho_1 \rho_2 \rho_3} \phi^l_{\rho_4 )} + \gamma^j_{\;ik} \phi^k_{\rho_1 \rho_2 \rho_3 \rho_4} \, , \\
    &N^{j \mu_\mathsf{S}}_{\rho_\mathsf{R} i} &&= \binom{\mathsf{r}}{\mathsf{s}} N^j_{(\rho_{\mathsf{R} - \mathsf{S}}| i} \delta^{\mu_\mathsf{S}}_{|\ldots \rho_\mathsf{r})} \, . \tag{\stepcounter{equation}\theequation}
\end{alignat*}

\subsection*{Applying the All-order Covariant Geometry to Amplitudes}

In the main text, we developed a prescription for expressing tree-level scattering amplitudes of a four-derivative $\mathsf{q}=1$ Lagrangian covariantly in terms of d-tensors. Here we demonstrate how to derive covariant expressions for tree-level scattering amplitudes of the most general four-derivative Lagrangian (which sits on the $\mathsf{q}=4$ jet bundle):
\begin{multline}
    \mathcal{L} = V(\phi^a) + \eta^{\mu\nu} [ g_{ij}(\phi^a) \, \phi^i_\mu \phi^j_\nu + h_i(\phi^a) \, \phi^i_{\mu\nu} ] + \eta^{\mu\nu} \eta^{\rho\sigma} [ c_i(\phi^a) \, \phi^i_{\mu\nu\rho\sigma} + c_{ij}(\phi^a) \, \phi^i_{\mu\nu\rho} \phi^j_\sigma + d_{ij}(\phi^a) \, \phi^i_{\mu\nu} \phi^j_{\rho\sigma} + e_{ij}(\phi^a) \, \phi^i_{\mu\rho} \phi^j_{\nu\sigma} \\
    + c_{ijk}(\phi^a) \, \phi^i_{\mu\nu} \phi^j_\rho \phi^k_\sigma + d_{ijk}(\phi^a) \, \phi^i_{\mu\rho} \phi^j_\nu \phi^k_\sigma + c_{ijkl}(\phi^a) \, \phi^i_\mu \phi^j_\nu \phi^k_\rho \phi^l_\sigma ] \, ,
\end{multline}
using the connection $\grad$ in \cref{eq:allorderNlinear};  any more complicated theory can be handled in the same way. Without loss of generality, the Wilson coefficients are assumed to be appropriately symmetric, satisfying $c_{ijkl} = c_{klij}$ in particular. The potential and kinetic terms have already been treated on $J^2(\pi)$ and the results carry over exactly, so we focus on the four-derivative terms.

The recipe goes as follows:
\begin{enumerate}
    \item As a function of kinematic variables, the scattering amplitude is specified by partial derivatives of Wilson coefficients $c$ evaluated at the vacuum on $\mathcal{M}$, resulting from the Feynman rules.
    \item Our goal is to replace each partial derivative with a d-tensor derived from $\mathcal{L}$ using $\grad$ in some well-chosen coordinates, because the invariance of the amplitude will then imply that the total d-tensor expression holds in any coordinates. The most convenient choice is the normal coordinates of $\nabla$. Any intermediate expression that holds only in these coordinates will be indicated with $\rightarrow$.
    \item First working in general coordinates, take the appropriate v-covariant derivatives on $\mathcal{L}$ and restrict to the null section to pick out a d-tensor $T(\phi^a, \phi^a_\Lambda = 0)$ that matches a tensor $t(\phi^a)$ on $\mathcal{M}$ containing $c(\phi^a)$, possibly with extra terms involving $\gamma$ from the non-linear connection $N$ and other Wilson coefficients.
    \item If there is no extra term, invoke \cref{eq:normalreplace} and \cref{eq:ttoT} to achieve our goal for partial derivatives of any order on $c$ at the vacuum. This step requires evaluating at $\phi^a = \bar{\phi}^a$ in normal coordinates.
    \item Otherwise, take the appropriate number of h-covariant derivatives on $T(\phi^a, \phi^a_\Lambda = 0)$ and evaluate at $\phi^a = \bar{\phi}^a$ in normal coordinates before attempting to cancel the extra terms. Any $\gamma$ at the vacuum can be rewritten using the curvature $R$ of $\nabla$ and hence the hh-curvature d-tensor $\mathcal{R}$ of $\grad$.
\end{enumerate}

Suppose for simplicity that we are interested in scattering up to four points. Then the following replacements for $\gamma$ suffice:
\begin{equation}
    \bar{\gamma}^i_{\;jk} \rightarrow 0, \quad \bar{\gamma}^i_{\;jk,l} \rightarrow - \frac{2}{3} \bar{R}^i_{\;(jk)l}, \quad \bar{\gamma}^i_{\;jk,lm} \rightarrow \frac{1}{6} \bar{R}^i_{\;(lm)(j,k)} - \frac{5}{6} \bar{R}^i_{\;(jk)(l,m)} \, .
\end{equation}

The most straightforward Wilson coefficient to pick out, with no extra term on the null section, is
\begin{equation}
    \frac{1}{28} \eta_{\mu\nu} \eta_{\rho\sigma} \frac{\delta \mathcal{L}}{\delta \phi^i_{\mu\nu\rho\sigma}} \Bigr |_{\phi^a_\Lambda = 0} = c_i \, .
\end{equation}
\cref{eq:normalreplace,eq:ttoT} then generate its d-tensor counterpart at the vacuum for any partial derivative:
\begin{equation}
    \frac{1}{28} \eta_{\mu\nu} \eta_{\rho\sigma} \left ( \grad_{(k_1} \ldots \grad_{k_\mathtt{n})} \frac{\delta \mathcal{L}}{\delta \phi^i_{\mu\nu\rho\sigma}} \right ) \Bigr |_{(\phi^a, \phi^a_\Lambda) = (\bar{\phi}^a,\, 0)} + \mathcal{O} ( \mathcal{R} ) \rightarrow \bar{c}_{i,k_1 \ldots k_\mathtt{n}} \, .
\end{equation}

Next, there is an extra term in
\begin{equation}
    \frac{1}{16} \eta_{\mu\nu} \eta_{\rho\sigma} \frac{\delta^2 \mathcal{L}}{\delta \phi^i_{\mu\nu\rho} \, \delta \phi^j_\sigma} \Bigr |_{\phi^a_\Lambda = 0} = c_{ij} - 4 \gamma^k_{\;ij} c_k \, ,
\end{equation}
from which we must produce $\bar{c}_{ij}$, $\bar{c}_{ij,k}$ and $\bar{c}_{ij,kl}$. Higher-order derivatives are not needed at three or four points. Taking h-covariant derivatives at the vacuum in normal coordinates, we find
\begin{align}
    \frac{1}{16} \eta_{\mu\nu} \eta_{\rho\sigma} \frac{\delta^2 \mathcal{L}}{\delta \phi^i_{\mu\nu\rho} \, \delta \phi^j_\sigma} \Bigr |_{(\phi^a, \phi^a_\Lambda) = (\bar{\phi}^a,\, 0)} &\rightarrow \bar{c}_{ij} \, , \\
    \frac{1}{16} \eta_{\mu\nu} \eta_{\rho\sigma} \grad_k \frac{\delta^2 \mathcal{L}}{\delta \phi^i_{\mu\nu\rho} \, \delta \phi^j_\sigma} \Bigr |_{(\phi^a, \phi^a_\Lambda) = (\bar{\phi}^a,\, 0)} &\rightarrow \bar{c}_{ij,k} + \frac{8}{3} \bar{R}^l_{\;(ij)k} \, \bar{c}_l \, , \label{eq:cijk} \\
    \frac{1}{16} \eta_{\mu\nu} \eta_{\rho\sigma} \grad_l \grad_k \frac{\delta^2 \mathcal{L}}{\delta \phi^i_{\mu\nu\rho} \, \delta \phi^j_\sigma} \Bigr |_{(\phi^a, \phi^a_\Lambda) = (\bar{\phi}^a,\, 0)} &\rightarrow \bar{c}_{ij,kl} + \frac{2}{3} \bar{R}^m_{\;(ik)l} \, \bar{c}_{mj} + \frac{2}{3} \bar{R}^m_{\;(jk)l} \, \bar{c}_{im} \label{eq:cijkl} \\
    &\quad + \left ( \frac{10}{3} \bar{R}^m_{\;(ij)(k,l)} - \frac{2}{3} \bar{R}^m_{\;(kl)(i,j)} \right ) \bar{c}_m + \frac{8}{3} \bar{R}^m_{\;(ij)k} \, \bar{c}_{m,l} + \frac{8}{3} \bar{R}^m_{\;(ij)l} \, \bar{c}_{m,k} \, . \nonumber
\end{align}
To isolate $\bar{c}_{ij,k}$, we subtract the following from \cref{eq:cijk}:
\begin{equation}
    \frac{2}{21} \bar{\mathcal{R}}^l_{\;(ij)k} \, \eta_{\mu\nu} \eta_{\rho\sigma} \frac{\delta \mathcal{L}}{\delta \phi^l_{\mu\nu\rho\sigma}} \Bigr |_{(\phi^a, \phi^a_\Lambda) = (\bar{\phi}^a,\, 0)} = \frac{8}{3} \bar{R}^l_{\;(ij)k} \, \bar{c}_l \, .
\end{equation}
Similarly, noting that $\nabla_l \bar{c}_{m} \rightarrow \bar{c}_{m,l}$ and $\nabla_l \bar{R}^m_{\;ijk} \rightarrow \bar{R}^m_{\;ijk,l}$, the next partial derivative $\bar{c}_{ij,kl}$ can be isolated from \cref{eq:cijkl} by subtracting
\begin{equation}
\begin{gathered}
    \frac{1}{24} \mathcal{R}^m_{\;(ik)l} \, \eta_{\mu\nu} \eta_{\rho\sigma} \frac{\delta^2 \mathcal{L}}{\delta \phi^m_{\mu\nu\rho} \, \delta \phi^j_\sigma} + \frac{1}{24} \mathcal{R}^m_{\;(jk)l} \, \eta_{\mu\nu} \eta_{\rho\sigma} \frac{\delta^2 \mathcal{L}}{\delta \phi^i_{\mu\nu\rho} \, \delta \phi^m_\sigma} \\
    + \frac{1}{42} \left ( 5 \mathcal{R}^m_{\;(ij)(k/l)} - \mathcal{R}^m_{\;(kl)(i/j)} \right ) \eta_{\mu\nu} \eta_{\rho\sigma} \frac{\delta \mathcal{L}}{\delta \phi^m_{\mu\nu\rho\sigma}} + \frac{2}{21} \mathcal{R}^m_{\;(ij)k} \, \eta_{\mu\nu} \eta_{\rho\sigma} \grad_l \frac{\delta \mathcal{L}}{\delta \phi^m_{\mu\nu\rho\sigma}} + \frac{2}{21} \mathcal{R}^m_{\;(ij)l} \, \eta_{\mu\nu} \eta_{\rho\sigma} \grad_k \frac{\delta \mathcal{L}}{\delta \phi^m_{\mu\nu\rho\sigma}} \, ,
\end{gathered}
\end{equation}
evaluated at the vacuum $(\phi^a, \phi^a_\Lambda) = (\bar{\phi}^a,\, 0)$.

The two subsequent Wilson coefficients come in pairs:
\begin{align}
    \frac{1}{8} \eta_{\mu\nu} \eta_{\rho\sigma} \frac{\delta^2 \mathcal{L}}{\delta \phi^i_{\mu\nu} \, \delta \phi^j_{\rho\sigma}} \Bigr |_{\phi^a_\Lambda = 0} &= 4 d_{ij} + e_{ij} - 6 \gamma^k_{\;ij} c_k \, , \\
    \frac{1}{8} \eta_{\mu\rho} \eta_{\nu\sigma} \frac{\delta^2 \mathcal{L}}{\delta \phi^i_{\mu\nu} \, \delta \phi^j_{\rho\sigma}} \Bigr |_{\phi^a_\Lambda = 0} = \frac{1}{8} \eta_{\mu\sigma} \eta_{\nu\rho} \frac{\delta^2 \mathcal{L}}{\delta \phi^i_{\mu\nu} \, \delta \phi^j_{\rho\sigma}} \Bigr |_{\phi^a_\Lambda = 0} &= d_{ij} + 7 e_{ij} - 15 \gamma^k_{\;ij} c_k \, .
\end{align}
These arise from different contractions of the spacetime indices. We can eliminate one coefficient for the other:
\begin{align}
    \frac{1}{216} \left ( 7 \eta_{\mu\nu} \eta_{\rho\sigma} - \eta_{\mu\rho} \eta_{\nu\sigma} \right ) \frac{\delta^2 \mathcal{L}}{\delta \phi^i_{\mu\nu} \, \delta \phi^j_{\rho\sigma}} \Bigr |_{\phi^a_\Lambda = 0} &= d_{ij} - \gamma^k_{\;ij} c_k \, , \\
    \frac{1}{216} \left ( - \eta_{\mu\nu} \eta_{\rho\sigma} + 4 \eta_{\mu\rho} \eta_{\nu\sigma} \right ) \frac{\delta^2 \mathcal{L}}{\delta \phi^i_{\mu\nu} \, \delta \phi^j_{\rho\sigma}} \Bigr |_{\phi^a_\Lambda = 0} &= e_{ij} - 2 \gamma^k_{\;ij} c_k \, ,
\end{align}
and repeat the preceding exercise to derive d-tensor expressions for $\bar{d}_{ij}$, $\bar{d}_{ij,k}$, $\bar{d}_{ij,kl}$, $\bar{e}_{ij}$, $\bar{e}_{ij,k}$ and $\bar{e}_{ij,kl}$.

Moving on to three v-covariant derivatives, we have another pair:
\begin{align}
    \frac{1}{4} \eta_{\mu\nu} \eta_{\rho\sigma} \frac{\delta^3 \mathcal{L}}{\delta \phi^i_{\mu\nu} \, \delta \phi^j_\rho \, \delta \phi^k_\sigma} \Bigr |_{\phi^a_\Lambda = 0} &= 8 c_{ijk} + 2 d_{ijk} - 6 \gamma^l_{\;ik} c_{lj} - 6 \gamma^l_{\;ij} c_{lk} - 16 \gamma^l_{\;jk} d_{il} - 4 \gamma^l_{\;jk} e_{il} \nonumber \\
    &\quad - 8 \left ( 2 \gamma^l_{\;ik,j} + \gamma^l_{\;jk,i} - 2 \gamma^l_{\;mi} \gamma^m_{\;jk} - \gamma^l_{\;mj} \gamma^m_{\;ik} - 3 \gamma^l_{\;mk} \gamma^m_{\;ij} \right ) c_l \, , \\
    \frac{1}{4} \eta_{\mu\rho} \eta_{\nu\sigma} \frac{\delta^3 \mathcal{L}}{\delta \phi^i_{\mu\nu} \, \delta \phi^j_\rho \, \delta \phi^k_\sigma} \Bigr |_{\phi^a_\Lambda = 0} &= \frac{1}{4} \eta_{\mu\sigma} \eta_{\nu\rho} \frac{\delta^3 \mathcal{L}}{\delta \phi^i_{\mu\nu} \, \delta \phi^j_\rho \, \delta \phi^k_\sigma} \Bigr |_{\phi^a_\Lambda = 0} \\
    &= 2 c_{ijk} + 8 d_{ijk} - 9 \gamma^l_{\;ik} c_{lj} - 9 \gamma^l_{\;ij} c_{lk} - 4 \gamma^l_{\;jk} d_{il} - 16 \gamma^l_{\;jk} e_{il} \nonumber \\
    &\quad - 12 \left ( 2 \gamma^l_{\;ik,j} + \gamma^l_{\;jk,i} - 2 \gamma^l_{\;mi} \gamma^m_{\;jk} - \gamma^l_{\;mj} \gamma^m_{\;ik} - 3 \gamma^l_{\;mk} \gamma^m_{\;ij} \right ) c_l \, .
\end{align}
Again, we can eliminate $d_{ijk}$ for $c_{ijk}$ or vice versa:
\begin{align}
    \frac{1}{20} \left ( 4 \eta_{\mu\nu} \eta_{\rho\sigma} - \eta_{\mu\rho} \eta_{\nu\sigma} \right ) \frac{\delta^3 \mathcal{L}}{\delta \phi^i_{\mu\nu} \, \delta \phi^j_\rho \, \delta \phi^k_\sigma} \Bigr |_{\phi^a_\Lambda = 0} &= 6 c_{ijk} - 3 \gamma^l_{\;ik} c_{lj} - 3 \gamma^l_{\;ij} c_{lk} - 12 \gamma^l_{\;jk} d_{il} \nonumber \\
    &\quad - 4 \left ( 2 \gamma^l_{\;ik,j} + \gamma^l_{\;jk,i} - 2 \gamma^l_{\;mi} \gamma^m_{\;jk} - \gamma^l_{\;mj} \gamma^m_{\;ik} - 3 \gamma^l_{\;mk} \gamma^m_{\;ij} \right ) c_l \, , \\
    \frac{1}{20} \left ( - \eta_{\mu\nu} \eta_{\rho\sigma} + 4 \eta_{\mu\rho} \eta_{\nu\sigma} \right ) \frac{\delta^3 \mathcal{L}}{\delta \phi^i_{\mu\nu} \, \delta \phi^j_\rho \, \delta \phi^k_\sigma} \Bigr |_{\phi^a_\Lambda = 0} &= 6 d_{ijk} - 6 \gamma^l_{\;ik} c_{lj} - 6 \gamma^l_{\;ij} c_{lk} - 12 \gamma^l_{\;jk} e_{il} \nonumber \\
    &\quad - 8 \left ( 2 \gamma^l_{\;ik,j} + \gamma^l_{\;jk,i} - 2 \gamma^l_{\;mi} \gamma^m_{\;jk} - \gamma^l_{\;mj} \gamma^m_{\;ik} - 3 \gamma^l_{\;mk} \gamma^m_{\;ij} \right ) c_l \, ,
\end{align}
and then take h-covariant derivatives at the vacuum in normal coordinates as before, yielding d-tensor counterparts to $\bar{c}_{ijk}$, $\bar{c}_{ijk,l}$, $\bar{d}_{ijk}$ and $\bar{d}_{ijk,l}$.

The last Wilson coefficient at four points is contained in
\begin{equation}
\begin{aligned}
    \frac{1}{16} \eta_{\mu\nu} \eta_{\rho\sigma} \frac{\delta^4 \mathcal{L}}{\delta \phi^i_\mu \, \delta \phi^j_\nu \, \delta \phi^k_\rho \, \delta \phi^l_\sigma} \Bigr |_{(\phi^a, \phi^a_\Lambda) = (\bar{\phi}^a,\, 0)} &\rightarrow 8 \bar{c}_{ijkl} + 2 \bar{c}_{ikjl} + 2 \bar{c}_{iljk} - 12 \bar{\gamma}^m_{\;l(i,jk)} \bar{c}_m \\
    &\quad - 3 ( \bar{\gamma}^m_{\;l(j,k)} \bar{c}_{mi} + \bar{\gamma}^m_{\;l(i,k)} \bar{c}_{mj} + \bar{\gamma}^m_{\;l(i,j)} \bar{c}_{mk} + \bar{\gamma}^m_{\;k(i,j)} \bar{c}_{ml} ) \, .
\end{aligned}
\end{equation}
The two other contraction structures $\eta_{\mu\rho} \eta_{\nu\sigma}$ and $\eta_{\mu\sigma} \eta_{\nu\rho}$ differ on the right hand side by only the first three terms, with coefficients $(2,8,2)$ and $(2,2,8)$ in that order. Taking the combination $5\eta_{\mu\nu} \eta_{\rho\sigma} - \eta_{\mu\rho} \eta_{\nu\sigma} - \eta_{\mu\sigma} \eta_{\nu\rho}$ leaves $(36,0,0)$, from which we can again derive a d-tensor expression for $\bar{c}_{ijkl}$ in normal coordinates. All the Wilson coefficients and their partial derivatives that show up in the three- and four-point amplitudes have now been written covariantly in normal coordinates, and the resulting expressions for the total amplitudes will hold in all coordinates. Extending the procedure to higher points is straightforward.

\subsection*{Multi-time Lagrange Spaces}

If we restrict to a special class of scalar field theories, the minimal geometry necessary to accommodate a Lagrangian $\mathcal{L}$ is actually not the jet bundle $J^\mathsf{q}(\pi)$, but the tangent bundle $T\mathcal{M}$. Assuming that only first field derivatives and fully symmetric Wilson coefficients appear in $\mathcal{L}$, e.g. $c_{ijkl} = c_{(ijkl)}$ in \cref{eq:4derivL}, we can embed it as a function on $T\mathcal{M}$ by identifying
\begin{equation}
    \eta^{\mu\nu} \phi^i_\mu \phi^j_\nu \rightarrow \varphi^i \varphi^j \, ,
\end{equation}
where $(\phi^i, \varphi^j)$ are coordinates that chart out $T\mathcal{M}$ as explained in the introduction. The Lagrangian turns the tangent bundle into what is called a Lagrange space. Instead of assuming a field space connection $\nabla$, one can use the fundamental tensor on $T\mathcal{M}$:
\begin{equation}
    g_{ij}(\phi^a, \varphi^b) \equiv \frac{1}{2} \frac{\partial^2 \mathcal{L}}{\partial \varphi^i \partial \varphi^j} \, ,
\end{equation}
and the coefficients $G_i$ from the Euler-Lagrange equations:
\begin{equation}
    g_{ij} \frac{d^2 \phi^j}{dx^2} + 2 G_i = 0, \quad G_i \equiv \frac{1}{4} \left ( \frac{\partial^2 \mathcal{L}}{\partial \varphi^i \partial \phi^l} \varphi^l - \frac{\partial \mathcal{L}}{\partial \phi^i} \right ) \, ,
\end{equation}
to construct a non-linear connection $N$ \cite{Miron:1994nvt}. Scattering amplitudes can then be covariantly expressed in terms of an $N$-linear connection $\grad$. But notably, detaching $N$ from $\nabla$ also leads to a vertical torsion component of $\grad$ that isolates higher-order Wilson coefficients, so that additional higher-derivative physics such as positivity bounds can be geometrically characterized \cite{Craig:2023wni}. The benefit of the Lagrange space framework is that the connections now contain physical information on the Lagrangian beyond the kinetic term, unlike jet bundles whose connections on their own have thus far offered no new information compared to the Riemannian geometry of $\mathcal{M}$.

Let us attempt to extend the geometry of Lagrange spaces to jet bundles, which would give rise to multi-time Lagrange spaces that embed higher-derivative physics in the connections. It suffices for our purposes to work at order $\mathsf{q}=1$. We seek to construct $N$ on $J^1(\pi)$ from the analogous fundamental tensor
\begin{equation}
    \mathfrak{g}^{\mu \nu}_{ij} \equiv \frac{1}{2} \frac{\partial^2 \mathcal{L}}{\partial \phi^i_\mu \partial \phi^j_\nu} \, ,
\end{equation}
and Euler-Lagrange equations
\begin{equation}
    \mathfrak{g}^{\mu \nu}_{ij} \frac{\partial^2 \phi^j}{\partial x^\mu \partial x^\nu} + 2 \mathfrak{G}_i = 0, \quad \mathfrak{G}_i \equiv \frac{1}{4} \left ( \frac{\partial^2 \mathcal{L}}{\partial \phi^i_\mu \partial \phi^l} \phi^l_\mu - \frac{\partial \mathcal{L}}{\partial \phi^i} \right ) \, .
\end{equation}
The coefficients $\mathfrak{G}_i$ obey the transformation law
\begin{equation}
\label{eq:Gtrans}
    \mathfrak{G}_l =  \tilde{\mathfrak{G}}_m \frac{\partial \tilde{\phi}^m}{\partial \phi^l} - \frac{\mathfrak{g}^{\mu\nu}_{kl}}{2} \frac{\partial^2 \phi^k}{\partial \tilde{\phi}^i \partial \tilde{\phi}^j} \tilde{\phi}^i_\mu \tilde{\phi}^j_\nu \, .
\end{equation}
Taking the d-tensor derivative $\partial / \partial \phi^n_\rho$ on \cref{eq:Gtrans} gets us to \cref{eq:N01trans} provided we can eliminate any field derivative dependence due to $\mathfrak{g}^{\mu\nu}_{kl}$ first. With one uncontracted index, the only obvious possibility is that $\mathfrak{g}^{\mu\nu}_{kl} = \eta^{\mu\nu} \mathsf{g}_{kl}$ is factorizable for some d-tensor $\mathsf{g}_{kl}$, so that $\mathsf{g}_{kl}$ can be eliminated before the differentiation by inverting it (where possible). If so, the following choice does the trick \cite{neagu2003}:
\begin{equation}
    N^j_{\rho i} = \eta_{\rho\sigma} \frac{\partial}{\partial \phi^j_\sigma} \left ( \mathsf{g}^{ik} \mathfrak{G}_k \right ) \, .
\end{equation}
As for $\grad$, we now have multiple choices:
\begin{subequations}
\begin{alignat}{2}
    F^i_{jk} &= \frac{\mathsf{g}^{il}}{2} \left( \frac{\delta \mathsf{g}_{lk}}{\delta \phi^j} + \frac{\delta \mathsf{g}_{lj}}{\delta \phi^k} - \frac{\delta \mathsf{g}_{jk}}{\delta \phi^l} \right) &&\text{ or } \, \frac{1}{\dim \mathcal{T}} \frac{\partial N^i_{\rho k}}{\partial \phi^j_\rho} \, , \\
    C^{i\rho}_{jk} &= \frac{\mathsf{g}^{il}}{2} \left(\frac{\partial \mathsf{g}_{lk}}{\partial \phi^j_\rho} + \frac{\partial \mathsf{g}_{lj}}{\partial \phi^k_\rho} - \frac{\partial \mathsf{g}_{jk}}{\partial \phi^l_\rho} \right) &&\text{ or } \, 0 \, .
\end{alignat}
\end{subequations}
This would have been an analogous covariant geometry to that of Lagrange spaces. But alas, it turns out that if $\dim \mathcal{T} > 1$, then the factorizability of $\mathfrak{g}^{\mu \nu}_{kl}$ implies that $\mathsf{g}_{kl}$ is independent of $\phi^a_\alpha$ and $\mathcal{L}$ must be a two-derivative Lagrangian; for a proof, see \cite{neagu2006}. Unsurprisingly, Lagrange spaces are recovered if $\mathcal{T}$ is one-dimensional, since $\phi^i_\mu$ can then be identified with $\varphi^i$ and $J^1(\pi)$ with $\mathbb{R} \times T\mathcal{M}$ \cite{balan2011}. However, the introduction of multi-dimensional spacetime is a fundamental hindrance to the covariant geometry of $J^1(\pi)$ that has yet to be solved.

The next best thing, which we proposed in the $\mathsf{q}=2$ analysis, is to assume the Levi-Civita connection of the kinetic term as $\nabla$ and adopt the jet bundle connections $N$ and $\grad$ that follow, so that the connections contain physical information from the Lagrangian. We note that the invariance of this choice of covariant jet bundle geometry under total derivatives precisely mirrors how a connection derived from the Euler-Lagrange equations would have behaved. While the connections themselves are then essentially no more descriptive than that in Riemannian geometry, any other information can be extracted from the fully general Lagrangian, of any derivative order and Wilson coefficient structure, that we can now accommodate by enlarging the manifold from the tangent bundle to the jet bundle.

\end{document}